\documentclass[showpacs,pra,aps,superscriptaddress,twocolumn,floatfix]{revtex4-1}
\usepackage{float,amsmath,amssymb,mathrsfs}
\usepackage{graphicx}
\newfont{\ensmathquatorze}{msbm10 scaled 1400}
\newfont{\ensmathonze}{msbm10 scaled 1100}
\newfont{\ensmathdix}{msbm10}
\newfont{\ensmathneuf}{msbm10 scaled 833}
\newfont{\ensmathhuit}{msbm10 scaled 694}
\newfam\ensmathfam
\textfont\ensmathfam=\ensmathonze
\scriptfont\ensmathfam=\ensmathdix
\scriptscriptfont\ensmathfam=\ensmathhuit
\def\sss{\scriptscriptstyle}
\DeclareMathAlphabet{\mathpzc}{OT1}{pzc}{m}{it}
\newcommand{\m}{\mathpzc{m}}
\def\ep{\mathrm{e}}
\allowdisplaybreaks[4]

\begin{document}

\title{Quantum fluctuations around black hole horizons in Bose-Einstein
  condensates}

\author{P.-\'E. Larr\'e} \affiliation{Univ. Paris Sud, CNRS,
  Laboratoire de Physique Th\'eorique et Mod\`eles Statistiques,
  UMR8626, F-91405 Orsay} \author{A. Recati} \affiliation{INO-CNR BEC
  Center and Dipartimento di Fisica, Universit\`a di Trento, via
  Sommarive 14, I-38123 Povo, Trento, Italy}\author{I. Carusotto}
\affiliation{INO-CNR BEC Center and Dipartimento di Fisica,
  Universit\`a di Trento, via Sommarive 14, I-38123 Povo, Trento,
  Italy}\author{N. Pavloff} \affiliation{Univ. Paris Sud, CNRS,
  Laboratoire de Physique Th\'eorique et Mod\`eles Statistiques,
  UMR8626, F-91405 Orsay}

\begin{abstract}

  We study several realistic configurations making it possible to realize an
  acoustic horizon in the flow of a one dimensional Bose-Einstein
  condensate. In each case we give an analytical description of the
  flow pattern, the spectrum of Hawking radiation and, the
  associated quantum fluctuations.  Our calculations confirm that the
  non local correlations of the density fluctuations previously
  studied in a simplified model provide a clear signature of Hawking
  radiation also in realistic configurations. In addition we
  explain by direct computation how this non local signal relates to
  short range modifications of the density correlations.

\end{abstract}

\maketitle

\section{Introduction}

During the last decade, it has been realized that Bose-Einstein
condensates (BECs) were promising candidates for producing acoustic
analogs of gravitational black holes, with possible experimental
signature of the elusive Hawking radiation. The acoustic analogy had
been proposed on a general setting by Unruh in 1981 \cite{Unruh}, and
its specific implementation using BECs has been first proposed by
Garay {\it et al.}, followed by many others \cite{HawkingBEC}. We are
now reaching a stage where experimental realizations and the study of
these systems are possible \cite{Jeff10} and it is important to
propose realistic configurations of acoustic black holes and possible
signatures of Hawking radiation in BECs. Other experimental
routes for
observing analog Hawking radiation effects are based on non linear
optical devices \cite{Phi08}
or surface waves on moving fluids \cite{Rou10}. Note that this last
option is restricted to the {\it stimulated regime}
where the Hawking radiation results from a disturbance external to the
system.

In this line, density correlations have been proposed in
Ref. \cite{correlations} as a tool making it possible to identify the {\it
  spontaneous} Hawking signal and to extract it from thermal
noise. The physical picture behind this idea is the same as the one
initially proposed by Hawking \cite{Hawking,Par04}: quantum
fluctuations can be viewed as constant emission and re-absorption of
virtual particles. These particles can tunnel out near the event
horizon and are then separated by the background flow (which is
subsonic outside the acoustic black hole and supersonic inside),
giving rise to correlated currents emitted away from the region of the
horizon.  In contrast to the gravitational case, the experimentalist
is able to extract information from the interior of an acoustic black
hole. It is thus possible to get insight on the Hawking effect by
measuring a correlation signal between the currents emitted inside
and outside the black hole. This two-body correlation
signal appears to be poorly affected by the thermal noise and seems to
be a more efficient measure of the Hawking effect than the direct
detection of Hawking phonons (see Ref. \cite{Rec09}).

In one dimensional (1D) flows of BECs, following a suggestion by
Leonhardt {\it et al.} \cite{Leo03}, it is possible, within a
Bogoliubov treatment of quantum fluctuations, to give a detailed
account of the one and two-body Hawking signals. This idea was fully
developed in Refs. \cite{Rec09} and \cite{Mac09} to obtain physical
predictions for specific configurations. Ref. \cite{Rec09} focused on
a schematic black hole configuration introduced in \cite{correlations}
and denoted as a ``flat profile configuration'' in the following: it
consists of a uniform flow of a 1D BEC in which the two-body
interaction is spatially modulated in order to locally modify the
speed of sound in the system -- forming a subsonic upstream region and
a supersonic downstream one -- although the velocity and the density
of the flow remain constant. However, this type of flow, with a
position dependent two-body interaction allowing an easy theoretical
treatment, is only possible in presence of an external potential
specially tailored so that the local chemical potential remains
constant everywhere (see details in Sec. \ref{flat}). This makes the
whole system quite difficult to realize experimentally. 

In the present work we propose simpler sonic analogs of
black holes for which a fully analytic theoretical treatment of the
quantum correlations is still possible. We present a detailed account
of the Bogoliubov treatment of quantum fluctuations in these settings
and show that density correlations provide, also in these realistic
configurations, a good evidence of the Hawking effect.  We also
discuss the recent work of Franchini and Kravtsov \cite{Fra09} who
proposed an interesting scenario for explaining the peculiarities of
the two-body density matrix $g^{(2)}$ in presence of an
horizon. Elaborating on the similarities of $g^{(2)}$ with the level
correlation function of non standard ensembles of random matrices
\cite{Can95}, one can argue that the non local features of $g^{(2)}$
typical for Hawking radiation should be connected to a modification of
its short range behavior. We spend some time for precisely discussing
this point in the framework of the Bogoliubov description of the
fluctuations. Our analytical study of the wave functions of the
excitations makes it possible to obtain an non-ambiguous confirmation of this
hypothesis.

The paper is organized as follows. In Sec. \ref{configs} we present
three configurations allowing to realize an acoustic horizon in a 1D BEC.
Then, in Sec. \ref{BdG-equations}, we discuss the practical
implementation of the Bogoliubov approach to these non uniform
systems. It appears convenient to describe the behavior of the
excitations in the system in terms of a $S$-matrix, the properties of
which are discussed in detail. This allows to describe the system
using an approach valid for all possible black hole configurations.
Within this framework, we study in Sec. \ref{radiation} the energy
current associated to the Hawking effect and in Sec. \ref{correl} the
density fluctuations pattern, putting special emphasis on its non
local aspects. As discussed above we consider in detail their
connection to short range modifications of the correlations. Finally
we present our conclusions in Sec. \ref{conclu}. Some technical points
are given in the appendices. In Appendix \ref{AppA} we present the low
energy behavior of the components of the $S$-matrix, in Appendix
\ref{AppC} we derive an expression for the energy current associated to
the Hawking radiation and in Appendix \ref{AppB} we precisely check
that the two-body density matrix fulfills a sum rule connecting the
short and long range behavior of the correlations in the system.

\section{The different black hole configurations}\label{configs}

We work in a regime which has been denoted as ``1D mean field'' in
Ref. \cite{Men02}. In this regime the system is described by a 1D
Heisenberg field operator $\hat{\Psi}(x,t)$, solution of the
Gross-Pitaevskii field equation. Writing
$\hat{\Psi}(x,t)=\hat{\Phi}(x,t)\exp(-{\rm i}\mu t/\hbar)$ this reads
\begin{equation}\label{e0}
{\rm i}\hbar\,\partial_t\hat{\Phi}=
-\frac{\hbar^2}{2m}\partial^2_x\hat{\Phi}+
[U(x)+g\hat{n}-\mu]\hat{\Phi}.
\end{equation}
In Eq. (\ref{e0}) $\mu$ is the chemical potential, fixed by boundary
conditions at infinity; $\hat{n}(x,t)=\hat{\Phi}^\dagger\hat{\Phi}$ is
the density operator and $U(x)$ is an external potential (its precise
form depends on the black hole configuration considered). $g$ is a non
linear parameter which depends on the two-body interaction within the
BEC and on the transverse confinement. Both are possibly position
dependent. For a repulsive effective two-body interaction described by
a positive 3D $s$-wave scattering length $a$ and for a
transverse harmonic trapping of pulsation $\omega_\perp$, one has
$g=2a\hbar\omega_\perp$ \cite{Ols98}. In the flat profile
configuration of Ref. \cite{correlations} $g$ depends on the position
$x$ (see Sec. \ref{flat}), whereas it is constant in the realistic
configurations introduced below and respectively denoted as delta peak
(Sec. \ref{delta}) and waterfall (Sec. \ref{waterfall})
configurations.

Within the Bogoliubov approach, in the quasi-condensate regime, the
quantum field operator $\hat{\Phi}$ is se\-pa\-ra\-ted in a classical
contribution $\Phi$ describing the background flow pattern plus a small
quantum correction $\hat{\psi}$. In all the configurations we
consider, the flow pattern is stationary and one thus writes
\begin{equation}\label{e5b}
\hat{\Phi}(x,t)=\Phi(x)+\hat{\psi}(x,t),
\end{equation}
$\Phi(x)$ being the solution of the classical stationary Gross-Pitaevskii
equation
\begin{equation}\label{e0bis}
\mu\Phi=-\frac{\hbar^2}{2m}\partial^2_x\Phi+
[U(x)+g|\Phi|^2]\Phi.
\end{equation}
A black hole configuration corresponds to a disymmetry between the
upstream flow and the downstream one, separated by the event
horizon. In the following we use a subscript ``$u$'' for upstream and
``$d$'' for downstream. The downstream region corresponds to $x>0$ and is
supersonic. The upstream region corresponds to $x<0$ and
is subsonic (see, however, the remark at the end of Sec. \ref{waterfall}). 
We thus write
\begin{equation}\label{e1}
\Phi(x)=\left\{\begin{array}{lcl}
\sqrt{n_u}\exp({\rm i} k_u x)\phi_u(x) & \mbox{for} & x<0, \\
\sqrt{n_d}\exp({\rm i} k_d x)\phi_d(x) & \mbox{for} & x>0.
\end{array}\right.
\end{equation}
In (\ref{e1}) $\lim_{x\to-\infty}|\phi_u(x)|=1$ and
$\lim_{x\to+\infty}|\phi_d(x)|=1$, so that $n_u$ and $n_d$ are
respectively the upstream and downstream asymptotic densities. Also
$k_\alpha = m V_\alpha/\hbar$ ($\alpha=u$ or $d$), where $V_u$ is
the asymptotic upstream flow velocity and $V_d$ the asymptotic
downstream one ($V_u$ and $V_d$ are both positive).

In the following we denote the asymptotic velocities of sound as
$c_u$ and $c_d$ with $m c_\alpha^2=g_\alpha n_\alpha$, where
$g_{u,d}=\lim_{x\to-\infty,+\infty}g(x)$ (we keep the possibility of a
position dependent $g$ coefficient in order to treat the flat profile
configuration of Ref. \cite{correlations}). We also introduce the
healing lengths $\xi_\alpha=\hbar/(m c_\alpha)$ and the Mach numbers
$\m_\alpha=V_\alpha/c_\alpha$. In a black hole
configurations $\m_u<1$ and $\m_d>1$.

Denoting $U_{u,d}=\lim_{x\to -\infty,+\infty}U(x)$
one gets from (\ref{e0bis}) and (\ref{e1})
\begin{equation}\label{e01}
\frac{\hbar^2k^2_\alpha}{2m} + U_\alpha + g_\alpha n_\alpha = \mu
\quad\mbox{and}\quad
n_u V_u = n_d V_d.
\end{equation}
The first of these equations corresponds to the equality of the
asymptotic chemical potentials and is required for a stationary
flow; the second equation corresponds to current
conservation in a stationary flow.

The precise form of the flow pattern is specified by the functions
$\phi_u(x)$ and $\phi_d(x)$ which depend on the configuration
considered. In all the configurations treated below $\phi_d(x)$ is a
constant of the form
\begin{equation}\label{down}
\phi_d(x)=\exp({\rm i}
\beta_d),
\end{equation}
meaning that the downstream flow pattern is flat with a constant
density and velocity. The value of $\beta_d$ depends on the
configuration considered. As for the upstream flow pattern, the
stationary flow condition imposes $\lim_{x\to-\infty}\phi_u(x)=\exp({\rm i}
\beta_u)$, where $\beta_u$ is a constant.

After having defined the notations and the common aspects of all the
flow patterns, we now give the precise value of the
configuration-dependent parameters.

\subsection{Flat profile configuration}\label{flat}

We recall here the value of the parameters in the flat profile
configuration studied in \cite{correlations,Rec09}. In this case the
$\phi_\alpha$ functions of Eq. (\ref{e1}) assume a very simple value:
$\phi_u(x)=\phi_d(x)=1$ (and thus $\beta_u=\beta_d=0$). One has
\begin{equation}\label{gradU}
U(x)=\left\{\begin{array}{lcl}
U_u & \mbox{for} & x<0, \\
U_d & \mbox{for} & x>0,
\end{array}\right.
\end{equation}
and
\begin{equation}\label{gradg}
g(x)=\left\{\begin{array}{lcl}
g_u & \mbox{for} & x<0, \\
g_d & \mbox{for} & x>0,
\end{array}\right.
\end{equation}
chosen so that a flow with $V_u=V_d\equiv V_0$ and $n_u=n_d\equiv
n_0$ is solution of Eqs. (\ref{e0bis}) and (\ref{e01}); i.e.,
Eq. (\ref{e1}) reduces to $\Phi(x)=\sqrt{n_0}\exp({\rm i} k_0 x)$ for
all $x$ ($k_0 = m V_0/\hbar$).  This imposes
\begin{equation}\label{gradino1}
\frac{c_d}{c_u}=\frac{\m_u}{\m_d}=\frac{\xi_u}{\xi_d},
\end{equation}
and
\begin{equation}\label{gradino2}
g_u n_0 + U_u = g_d n_0 + U_d.
\end{equation}
We finally note that in the flat profile configuration one has
$c_d<V_d=V_u<c_u$.

In the numerical simulations of Refs. \cite{correlations,Mac09}, a
genera\-li\-za\-tion of this step-like configuration has been used; one
considers smooth $U(x)$ and $g(x)$ functions imposing the continuous
version of (\ref{gradino2}): $g(x)n_0+U(x)=\rm{C^{st}}$.  The
theoretical approach is the same as in Ref. \cite{Rec09} but the
Bogoliubov-de Gennes equations [Eq. (\ref{e7}) below] have to be
solved numerically whereas the step-like confi\-gu\-ra\-tion characterized
by Eqs. (\ref{gradU}) and (\ref{gradg}) allows for an analytical
treatment.

The flat profile configuration can be numerically implemented in a
dynamical way as explained in Ref. \cite{correlations}. However, it is
fair to say that the corresponding experiment seems rather difficult
to realize. Moreover the flat profile configuration is very sensitive
to the total atom number, a quantity which is not easily controlled
experimentally. Besides a local monitoring of $g(x)$ has not
yet been demonstrated. There are the reasons why in the following
subsections we introduce two new types of sonic horizon which can be
implemented experimentally more easily.

\subsection{Delta peak configuration}\label{delta}

\begin{figure}
\includegraphics*[width=0.99\linewidth]{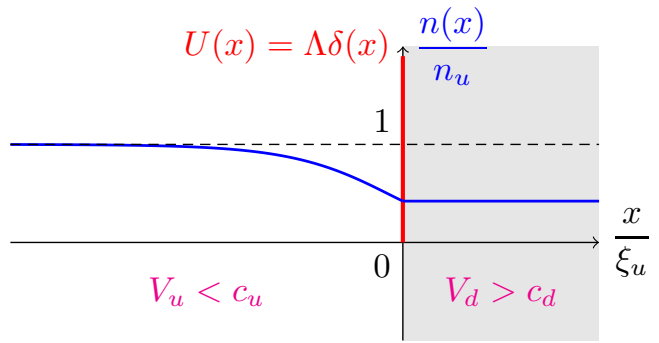}
\caption{(Color online) Density profile in the delta peak
  con\-fi\-gu\-ra\-tion. The flow is directed toward positive $x$. The
  delta potential is represented by a (red) vertical straight
  line. The density in the upstream region ($x<0$) is a portion of a
  dark soliton (see the text). The region $x>0$ is supersonic. It is
  shaded in the plot for recalling that it corresponds to the interior
  of the equivalent black hole. We keep this convention in
  Figs. \ref{fig2}, \ref{fig3} and \ref{fig5}.}
\label{fig1}
\end{figure}

In this configuration the non linear coefficient $g$ is constant and
the external potential is a repulsive delta peak: $U(x)=\Lambda
\delta(x)$, with $\Lambda>0$. It has been noticed in
Ref. \cite{Leb01} that one can find in this case a stationary profile
with a flow which is subsonic far upstream and supersonic downstream
(i.e., a black hole configuration). The upstream flow corresponds to a
portion of a dark soliton profile. More precisely, for $x<0$, one has
\begin{equation}\label{up}
\phi_u(x)=\cos\theta\,\tanh\left(\frac{x-x_0}{\xi_u}\cos\theta\right)-{\rm
i}\sin\theta,
\end{equation}
where $\sin\theta=\m_u$, and one can restrict oneself to
$\theta\in[0,\pi/2]$ (then $\beta_u=\pi+\theta$). As is also the case
for the other configurations studied in the present work, the
downstream flow has a constant density and velocity
[cf. Eq. (\ref{down})]. The typical profile is displayed in
Fig. \ref{fig1}.

Once $\m_u=V_u/c_u$ is fixed ($<1$) all the other
pa\-ra\-me\-ters of the flow are determined by Eqs. (\ref{e01}). Defining
$y=\frac{1}{2}\left(-1+\sqrt{1+8/\m_u^2}\right)$ one gets
\begin{equation}\label{e4}
\frac{n_u}{n_d}=\frac{V_d}{V_u}=y,\quad\frac{\m_d}{\m_u}=y^{\frac{3}{2}}
,\quad\frac{c_d}{c_u}=\frac{1}{\sqrt{y}}=\frac{\xi_u}{\xi_d}.
\end{equation}
By imposing continuity of the wave function
[$\Phi(0)=\sqrt{n_d}\exp({\rm i}\beta_d)= \sqrt{n_u} \,
\phi_u(0)$] and the appropriate mat\-ching of its first derivative
[$\partial_x\Phi(0^+)-\partial_x\Phi(0^-) = 2 m \hbar^{-2} \Lambda
\Phi(0)$] one gets
\begin{align}
\label{e5a1}
&\sin\beta_d=-\m_u\sqrt{y}, \\
\label{e5a2}
&\frac{x_0}{\xi_u}=\frac{1}{\cos\theta}
\tanh^{-1}\left(\sqrt{\frac{y-1}{2}}\tan\theta\right),
\end{align}
and also
\begin{equation}\label{e5bis}
\Lambda=\frac{\lambda\hbar^2}{m\xi_u}\quad\mbox{with}\quad
\lambda=\m_u\left(\frac{y-1}{2}\right)^{\frac{3}{2}}.
\end{equation}
In this configuration one has $V_u<c_d<c_u<V_d$ which corresponds
to a black hole type of horizon. Related work for a double barrier
configuration recently appeared in \cite{Zap11}.

Note that the flow depicted in Fig.~\ref{fig1} corresponds to a very specific
case in the parameter space spanned by the intensity of the delta
potential and the flow velocities, which is on the verge of becoming
time dependent (see Ref. \cite{Leb01}). This is reflected by the fact
that, for this configuration, $\m_u$ and $\m_d$ cannot be fixed
independently (in contrast to what occurs for the flat profile case).
One might thus legitimately expect to face a fine tuning problem to
expe\-ri\-men\-tally fulfill all the required boun\-da\-ry conditions
(\ref{e4}) (\ref{e5a1}), (\ref{e5a2}) and (\ref{e5bis}). One could
also argue that a delta potential is non standard and that the
specific structure of the flow pattern displayed in Fig. \ref{fig1}
would disappear for a more realistic potential. However, it is shown
in Ref. \cite{Car11} that this configuration can be rather easily
obtained by launching a 1D condensate on a localized obstacle (not
necessarily a delta peak). In this case, there exists a sizable range
of parameters where, after ejection of an upstream dispersive shock
wave, the long time flow pattern becomes of the type illustrated in
Fig. \ref{fig1}.

\subsection{Waterfall configuration}\label{waterfall}

\begin{figure}
\includegraphics*[width=0.99\linewidth]{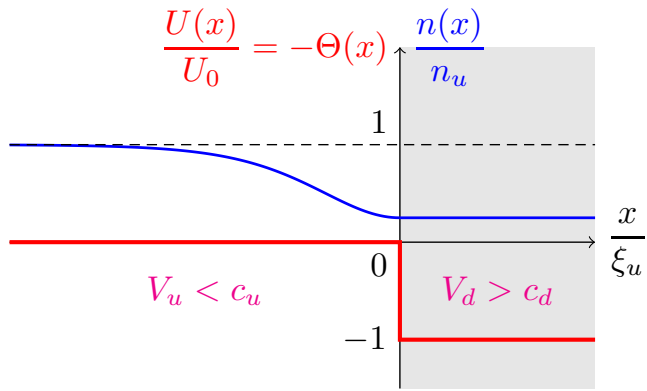}
\caption{(Color online) Same as Fig. \ref{fig1} for the
waterfall configuration.}
\label{fig2}
\end{figure}

In this configuration the two-body interaction is cons\-tant and the
external potential is a step function of the form $U(x)=-U_0
\Theta(x)$, where $\Theta$ is the Heaviside function (and $U_0>0$). In
this case, a stationary profile with a flow which is subsonic upstream
and supersonic downstream, i.e., a black hole configuration, has been
identified in Ref. \cite{Leb03}. The upstream profile is, as for the
delta peak configuration, of the form (\ref{up}), with here $x_0=0$,
i.e., the upstream profile is exactly one half of a dark soliton. The
corresponding density profile is displayed in Fig. \ref{fig2}.

The equalities (\ref{e01}) and the continuity of the order
parameter at the origin impose here
\begin{equation}\label{e2}
\frac{V_d}{V_u}=\frac{n_u}{n_d}=
\frac{1}{\m_u^2}=\m_d=\left(\frac{\xi_d}{\xi_u}\right)^2 
=\left(\frac{c_u}{c_d}\right)^2,
\end{equation}
$\exp({\rm i}\beta_d)=-{\rm i}$, and
\begin{equation}\label{e3}
\frac{U_0}{g n_u}=\frac{\m_u^2}{2}+\frac{1}{2\m_u^2}-1.
\end{equation}
In this configuration one has $V_u=c_d<c_u<V_d$
which corresponds to a black hole type of horizon.

The remark given at the end of Sec. \ref{delta} is here also in order:
there might be a fine tuning problem for verifying Eqs. (\ref{e2}) and
(\ref{e3}). Although we did not perform here the time dependent
analysis done in Ref. \cite{Car11} for the delta peak configuration, we
believe that, also in the present case, it is possible to dynamically
reach the stationary configuration depicted in Fig. \ref{fig2}. This
is supported by the experimental results presented by the Technion
group \cite{Jeff10} who studied a very similar configuration (with the
additional complication of the occurrence of a white hole
horizon). These results show no important time dependent features near
the black hole horizon and we are thus led to consider that the
stationary waterfall configuration of the type illustrated in
Fig. \ref{fig2} is stable and can be reached experimentally.

We make here a remark which is also relevant for the delta peak
configuration: the precise location of the sonic horizon is not well
defined. In both configurations (waterfall or delta peak) one may
define a local sound velocity, and the point where the local flow
velocity exceeds the local speed of sound can be chosen as the
location of the sonic horizon. Then one finds that the sonic horizon
is located slightly upstream the interface $x=0$ (in the waterfall
configuration for instance, at $x=0$ the local flow velocity is
already $\m_u^{-2}$ times larger than the local sound
velocity). However, the local sound velocity is an appro\-xi\-mate
concept, only rigorously valid in regimes where the BEC density varies
over typical length scales much larger than the healing length. This
is not the case in the waterfall and delta peak configurations near
$x=0$ and, as a result, the concept of sonic horizon is ill
defined. In the fully quantum treatment presented below we do not use
this concept: the important point for our analysis is simply that the
upstream flow velocity is asymptotically (i.e., when $x\to -\infty$) larger
than $c_u$. Hence, for preciseness we do not state that the
upstream flow is subsonic, but that it is {\it asymptotically}
subsonic.

\section{Fluctuations around the stationary
  profile}\label{BdG-equations}

In this section we establish a basis set in each of the flow regions
(upstream and downstream) which will be used in
Sec. \ref{quantization} for describing the quantum fluctuations in the
system. The simplest way to obtain this basis set is to start from
expression (\ref{e5b}), with $\Phi(x)$ given by (\ref{e1}), and to
treat $\hat{\psi}(x,t)$ as a small time dependent classical field,
denoted as $\psi(x,t)$ in the present section, with
$\Phi(x)+\psi(x,t)$ solution of the classical version of
(\ref{e0}). One looks for a normal mode of the form
\begin{equation}\label{e6}
\psi(x,t)=\ep^{{\rm i} k_\alpha x}
\left[\bar{u}_\alpha(x,\omega) \ep^{-{\rm i} \omega t}
+\bar{w}_\alpha^*(x,\omega) \ep^{{\rm i} \omega t}\right],
\end{equation}
with $\alpha=u$ for $x<0$ and $\alpha=d$ for $x>0$. $\Phi(x,t)$
defined by Eqs. (\ref{e5b}) and (\ref{e6}) describes small
oscillations with pulsation $\omega$ of the order parameter around the
ground state $\Phi(x)$. In the following we drop the $\omega$
dependence of functions $\bar{u}_\alpha$ and $\bar{w}_\alpha$ for
legibility. We also write $X_\alpha=x/\xi_\alpha$ (and then $k_\alpha
x=\m_\alpha X_\alpha$). Linearizing the Gross-Pitaevskii equation,
one gets at first order in $\psi$:
\begin{equation}\label{e7}
\varepsilon_\alpha
\begin{pmatrix}\bar{u}_\alpha \\ \bar{w}_\alpha\end{pmatrix}
=
\mathcal{L}_\alpha
\begin{pmatrix}
\bar{u}_\alpha \\ \bar{w}_\alpha
\end{pmatrix},
\end{equation}
with
\begin{equation}\label{lbogo}
\mathcal{L}_\alpha =
\begin{pmatrix}H_\alpha-{\rm i}\m_\alpha\partial_{X_\alpha} & \phi_\alpha^2 \\
-(\phi_\alpha^*)^2 & -H_\alpha-{\rm i}\m_\alpha\partial_{X_\alpha}
\end{pmatrix},
\end{equation}
where $\varepsilon_\alpha=\hbar\omega/(g_\alpha n_\alpha)$ and
$H_\alpha=-\frac{1}{2}\partial^2_{X_\alpha}+2|\phi_\alpha|^2-1$.
Hence, the column vector formed by $\bar{u}_\alpha$ and
$\bar{w}_\alpha$ is an eigen-vector of the so called Bogoliubov-de
Gennes Hamiltonian $\mathcal{L}_\alpha$.

The present section is organized as follows. We first consider
solutions of (\ref{e7}) for $X_\alpha\in\mathbb{R}$ in
Sec. \ref{general}.  We give the expression of these solutions in
sections \ref{downstream} and \ref{upstream}, specifying only
what we need for the following step: that is, for $\alpha=d$, we only
display the form of the solution when $x>0$, and for $\alpha=u$, we
only display the form of the solution when $x<0$. The most
general fluctuation of pulsation $\omega$ is a linear combination of
eigen-modes for the upstream region glued at $x=0$ with a linear
combination of the downstream eigen-modes. We explain how this
matching is done in Sec. \ref{matching}. Finally, in
Sec. \ref{scattering-modes}, we specify the form of the scattering
modes which are the appropriate modes used for quantizing the fluctuations
in Sec. \ref{quantization}.

\subsection{Properties of the eigen-functions of the
Bogoliubov-de Gennes equation}\label{general}

\begin{figure}
\includegraphics*[width=0.99\linewidth]{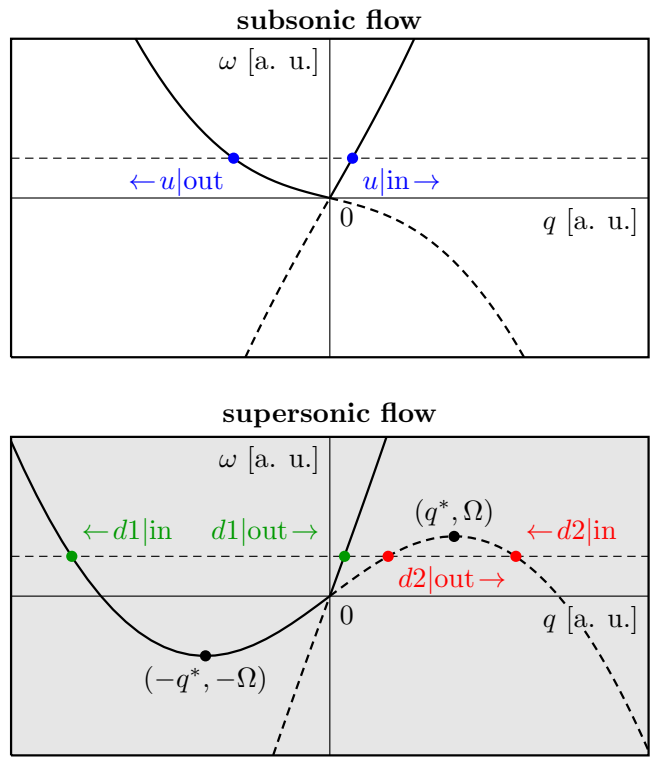}
\caption{(Color online) Dispersion relation (\ref{e8}). In each plot
  the horizontal dashed line is fixed by the chosen value of
  $\omega$. The $q_\ell(\omega)$'s are the corresponding
  abscissae. Only the real eigen-modes are represented; their
  denomination is explained in the text; their direction of
  propagation (left or right) is represented by an arrow. The part of
  the dispersion relation corresponding to negative norm states (see
  the text) is represented with a dashed line. The upper plot
  corresponds to a subsonic flow. The lower one corresponds to a
  supersonic flow; it is shaded in order to recall that it describes
  the situation inside the black hole.}\label{fig3}
\end{figure}

The relevant eigen-functions of (\ref{e7}) are of the form
\begin{equation}\label{e6b}
\begin{pmatrix}
\bar{u}_\ell(x) \\ \bar{w}_\ell(x)\end{pmatrix} =
\ep^{{\rm i} Q_\ell X_\alpha}
\begin{pmatrix}
{\cal U}_\ell(x) \\ {\cal W}_\ell(x)
\end{pmatrix},
\end{equation}
where the functions ${\cal U}_\ell(x)$ and ${\cal W}_\ell(x)$ are
constant for $|x|\to\infty$ (more precisely in the domain where
$\phi_\alpha$ is constant). Their exact form will be specified later
[Eqs. (\ref{e11}) and (\ref{e9})]. In Eq. (\ref{e6b})
the $Q_\ell$'s are the dimensionless wave vectors of the
Bogoliubov modes, solutions of
\begin{equation}\label{e8}
(\varepsilon_\alpha-\m_\alpha Q)^2=\omega_{\rm\sss B}^2(Q),
\end{equation}
where
\begin{equation}\label{bog}
\omega_{\rm\sss B}(Q)=Q\sqrt{1+\frac{Q^2}{4}}
\end{equation}
is the Bogoliubov dispersion relation in a condensate at rest (written
in dimensionless form). Note that $Q_\ell$ -- solution of (\ref{e8})
-- is sometimes complex; this fact is taken into account in the
following. In particular, for $\alpha=u$ ($\alpha=d$) one should
discard values of the wave vector such that ${\rm Im}(Q_\ell)>0$
(${\rm Im}(Q_\ell)<0$).  This corresponds to eliminating the
evanescent channels in the region where they are divergent. For
instance, a mode with $\mbox{Im}(Q_\ell)<0$ diverges when
$x\to+\infty$, which will correspond to the supersonic region
(labeled $d$) in the following, and we thus discard it.  The
dispersion relations and the different real wave vectors are displayed
in Fig. \ref{fig3}.

The index $\alpha$ of Eq. (\ref{e6}) is specified to $\ell$ in
(\ref{e6b}) for identifying the branch of the dispersion relation to
which the considered excitation pertains. For precise notations,
$\ell$ is taken as a double index, because it is clear from (\ref{e8})
that the values of the wave vectors are not the same in the subsonic
and supersonic regions, i.e., they depend on $\alpha$. Specifically,
when $\alpha=u$, $\ell\in\{u|{\rm in},u|{\rm out},u|{\rm
eva}\}$. When $\alpha=d$, there are two cases, depending if $\omega$
is lower or greater than a certain threshold $\Omega$. If
$\omega<\Omega$, $\ell\in\{d1|{\rm in},d1|{\rm out},d2|{\rm
in},d2|{\rm out}\}$ and when $\omega>\Omega$, $\ell\in\{d1|{\rm
in},d1|{\rm out},d|{\rm eva}\}$.

We have chosen to label the real eigen-modes as ``in'' (such as $d1|{\rm
in}$ for instance) or ``out'' (such as $u|{\rm out}$) depending if
their group velocity [its explicit expression is given below, Eq.
(\ref{e11b})] points toward the horizon (for the ``in'' modes) or
away from the horizon (for the ``out'' modes) in a black hole
configuration, i.e., with a subsonic region at left of the horizon,
the supersonic region being at the right. The wave vectors labeled
$u|{\rm eva}$ and $d|{\rm eva}$ are complex and correspond to
evanescent channels (as explained above, one selects the complex
$Q_{\ell}$'s which describe waves decaying at infinity).

The threshold $\Omega$ appearing in the lower plot of Fig. \ref{fig3}
is reached only for a supersonic flow, for a wave vector $q^*$ such that
\begin{equation}\label{t}
q^*\xi_{d}=Q^*_d=\left(-2+\frac{\m_d^2}{2}+
\frac{\m_d}{2}
\sqrt{8+\m_d^2}\right)^{\frac{1}{2}}.
\end{equation}
The existence of this threshold is a consequence of the behavior
of the large momentum part of the dispersion relation $\omega_{\rm\sss
B}(Q)$ of a condensate at rest. More precisely, the part of the
dispersion relation which shows a local maximum in the supersonic
region corresponds to the particular solution of (\ref{e8}) where
$\varepsilon_d-\m_d Q = -\omega_{\rm\sss B}(Q)$. At
$Q=Q^*_d$ one has exactly $\partial\omega_{\rm\sss B} / \partial
Q=\m_d$, and for $Q>Q^*_d$, one has $\partial\omega_{\rm\sss B}
/ \partial Q>\m_d$. Hence the part of the spectrum with $Q>Q^*_d$
corresponds to excitations whose group velocity in the frame where the
condensate is at rest ($\partial\omega_{\rm\sss B}/ \partial Q$) is
larger than the flow velocity $\m_d$ (we use here
dimensionless quantities). One can have $\partial\omega_{\rm\sss B}
/ \partial Q>\m_d >1$ only because the dispersion relation (\ref{bog})
grows faster than linear at large $Q$. We will see in
Sec. \ref{radiation} that the corresponding waves play an important
role in the zero temperature Hawking signal.

It is easy to verify that, if $(\bar{u}_\alpha,\bar{w}_\alpha)$ is a
solution of (\ref{e7}) associated to an eigen-energy
$\varepsilon_\alpha$, then $(\bar{w}_\alpha^*,\bar{u}_\alpha^*)$ is
also a solution of (\ref{e7}), now associated to eigen-energy
$-\varepsilon_\alpha$. Besides, from expression (\ref{e6}), one
sees that both solutions describe the same perturbation of the
condensate. As a result, one can always select eigen-modes
with po\-si\-ti\-ve values of $\varepsilon_\alpha=\hbar\omega/(g_\alpha
n_\alpha)$, and in all the following we chose $\omega\in\mathbb{R}^+$.

One may also notice that the above symmetry of the wave function is
associated to the normalization of the eigen-modes. For instance, it
is clear that the symmetry operation changes the sign of $|{\cal
  U}_\ell|^2-|{\cal W}_\ell|^2$ and one can show by simple algebraic
manipulations that, for real $Q_\ell$, the sign of $|{\cal
  U}_\ell|^2-|{\cal W}_\ell|^2$ is the same as the sign of
$\varepsilon_\alpha - \m_\alpha Q_\ell$. In the following we denote
the eigen-modes for which $|{\cal U}_\ell|^2-|{\cal W}_\ell|^2>0$ as
having a positive nor\-ma\-li\-za\-tion (for a recent discussion of
this point see see, e.g. Ref. \cite{Bar10}, as well as Refs.
\cite{BR86,Fet98}). In Fig. \ref{fig3} their dispersion relation
is represented with a solid line, whereas the eigen-modes with
negative normalization are represented with a dashed line. In
particular, in the supersonic region, the $d2|{\rm in}$ and $d2|{\rm
  out}$ channels have negative norm for $0<\omega<\Omega$.

It was shown in \cite{Dal96} that each eigen-vector of equation
(\ref{e7}) is associated with a conserved (i.e., $x$ independent)
current $J_\ell$. We show below (see Sec. \ref{radiation}) how $J_\ell$
relates to the energy current in the system. $J_\ell$ is zero for
complex $Q_\ell$ (evanescent waves do not carry any
current); this can be proven directly in our specific case, but we do not
display the proof here. For real $Q_\ell$ one gets
\begin{align}\label{cc}
\notag J_\ell  & =
c_\alpha \Big[ (Q_\ell + \m_\alpha) |{\cal U}_\ell|^2 +
(Q_\ell - \m_\alpha) |{\cal W}_\ell|^2 \Big] \\
& + c_\alpha {\rm Im} \left(
{\cal U}_\ell^* \partial_{X_\alpha} {\cal U}_\ell
+ {\cal W}_\ell^* \partial_{X_\alpha} {\cal W}_\ell \right).
\end{align}
Going back to dimensioned quantities and using the $\bar{u}_\ell$ and
$\bar{w}_\ell$ functions, this reads
\begin{equation}\label{cc4}
  J_\ell = \frac{\hbar}{2m}
  \Big[ \bar{u}^*_{\ell} (k_\alpha-{\rm i}\partial_x) \bar{u}_{\ell}
  -
  \bar{w}^*_{\ell} (k_\alpha+{\rm i}\partial_x) \bar{w}_{\ell} \Big] + {\rm c.c.},
\end{equation}
where ``c.c.'' stands for ``complex conjugate''.

\subsection{Downstream region: $x>0$}\label{downstream}

We recall that in this region the flow is supersonic and that
the eigen-vectors are labeled with an
index $\ell\in\{d1|{\rm in},d1|{\rm out},d2|{\rm in},d2|{\rm out}\}$
when $\omega<\Omega$ and $\ell\in\{d1|{\rm in},d1|{\rm out},d|{\rm
eva}\}$ when $\omega>\Omega$.

Here $\phi_d^2$ appearing in Eq. (\ref{lbogo}) does not depend on $x$
and is equal to $\exp(2{\rm i}\beta_d)$ \cite{remarque}. This
implies that the functions ${\cal U}_\ell$ and ${\cal W}_\ell$ are
also $x$ independent. One finds
\begin{equation}\label{e11}
\begin{pmatrix}
{\cal U}_{\ell} \\ {\cal W}_{\ell}
\end{pmatrix}
=\frac{1}{{\cal C}_\ell}
\begin{pmatrix}(Q_\ell^2/2+E_{\ell})\ep^{{\rm i}\beta_d} \\
(Q_\ell^2/2-{E}_{\ell})\ep^{-{\rm i}\beta_d}
\end{pmatrix},
\end{equation}
with ${E}_{\ell}=\varepsilon_d-\m_dQ_\ell$ and ${\cal C}_\ell$ is a
normalization constant which we always chose real and positive. The
corresponding current is easily evaluated using Eq. (\ref{cc}). For
real $Q_\ell$ one gets
\begin{equation}\label{e11a}
J_\ell = V_{g}(Q_{\ell})
(|{\cal U}_\ell|^2 - |{\cal W}_\ell|^2),
\end{equation}
where
\begin{equation}\label{e11b}
V_{g}(Q_\ell) = c_d
\frac{\partial \varepsilon_d}{\partial Q_\ell} =
\frac{\partial \omega}{\partial q_\ell}
\end{equation}
is the group velocity in the laboratory frame and $q_\ell=Q_\ell/\xi_\alpha$
with here $\alpha=d$, but Eqs. (\ref{e11a}) and (\ref{e11b}) are valid also
for $\alpha=u$.

A typical choice for the normalization constant is
${\cal C}_\ell=|2\,{\rm Re}({E}_{\ell}^*Q_\ell^2)|^{1/2}$. This ensures that
$|{\cal U}_{\ell}|^2-|{\cal W}_{\ell}|^2=\pm 1$. For our case, it is
more appropriate to multiply the previous expression of ${\cal C}_\ell$ by
$|V_g(Q_\ell)|^{1/2}$, so that for real $Q_\ell$ one has $J_\ell=\pm
1$. Hence we chose
\begin{equation}\label{normem1} 
{\cal C}_\ell=|2\,\mbox{Re}({E}_{\ell}^*Q_\ell^2)V_g(Q_\ell)|^{\frac{1}{2}},
\end{equation}
which implies
\begin{equation}\label{norme}
|{\cal U}_{\ell}|^2-|{\cal W}_{\ell}|^2=\frac{\pm 1}{|V_g(Q_\ell)|}.
\end{equation}
The modes for which the factor $+1$ appears in the above expression
are the positive norm modes previously discussed. The others are the
negative norm modes. With the normalization (\ref{norme}), from
Eq. (\ref{e11a}), one sees that $J_\ell=1$ either for non evanescent
modes of positive norm propagating to the right, or for non evanescent
modes of negative norm propagating to the left. In the other non
evanescent cases (modes of positive norm propagating to the left or
modes of negative norm propagating to the right) $J_\ell=-1$. More
precisely: $J_{d1|{\rm out}}=+1=J_{d2|{\rm in}}$ and $J_{d1|{\rm
    in}}=-1=J_{d2|{\rm out}}$. We will see below (Sec. \ref{radiation}) that
$\hbar\omega J_\ell$ is the energy current associated to mode $\ell$ and
thus negative norm modes can be interpreted as carrying negative energy.

\subsection{Upstream region: $x<0$}\label{upstream}

We recall that in this region the flow is asymptotically subsonic and
that $\ell\in\{u|{\rm in},u|{\rm out},u|{\rm eva}\}$. In the flat
profile configuration one has $\phi_u(X_u)=1$ and the functions ${\cal
  U}_\ell$ and ${\cal W}_\ell$ have the same form as the ones
displayed in the previous section. Hence in the remainder of the
present subsection we concentrate on the delta peak and waterfall
configurations where $\phi_u$ depends on $X_u$. In this case the
functions ${\cal U}_\ell$ and ${\cal W}_\ell$ have a more complicated
expression than in the downstream region (see, e.g., Appendix A of
Ref. \cite{Bil05}). Defining
$\chi(X_u)=\cos\theta\tanh[(X_u-X_0)\cos\theta)]$, where
$X_0=x_0/\xi_u$ (we recall that in the waterfall configuration
$x_0=0$), one gets
\begin{equation}\label{e9}
\begin{pmatrix}{\cal U}_\ell(x) \\ {\cal W}_\ell(x)\end{pmatrix} =
\frac{1}{{\cal D}_\ell}
\begin{pmatrix}
\left[Q_\ell/2+\varepsilon_u/Q_\ell+{\rm i}\chi(X_u)\right]^2 \\
\left[Q_\ell/2-\varepsilon_u/Q_\ell+{\rm i}\chi(X_u)\right]^2
\end{pmatrix},
\end{equation}
where ${\cal D}_\ell$ is an arbitrary constant, the value of which is
determined by the normalization (see below). The current (\ref{cc})
corresponding to ${\cal U}_\ell$ and ${\cal W}_\ell$ given in
(\ref{e9}) is most easily evaluated at $X_u\to -\infty$, i.e., in a
region where ${\cal U}_\ell$ and ${\cal W}_\ell$ become independent of
$X_u$. In this region one has
\begin{equation}\label{e9b}
\begin{pmatrix}{\cal U}_\ell \\ {\cal W}_\ell\end{pmatrix}
\underset{X_u \to -\infty}{\longrightarrow}
\frac{1}{{\cal D}_\ell}\begin{pmatrix}
\left[Q_\ell/2+\varepsilon_u/Q_\ell-{\rm i}\cos\theta\right]^2 \\
\left[Q_\ell/2-\varepsilon_u/Q_\ell-{\rm i}\cos\theta\right]^2
\end{pmatrix}.
\end{equation}
Since $\phi_u$ tends to a constant [$\exp({\rm i}\beta_u)$] when
$X_u\to -\infty$, one could, in this region, use for the Bogoliubov
modes an expression similar to Eq. (\ref{e11}) which is used in the
downstream domain. Indeed, it is difficult to see it from the above
formula, but we have checked that (\ref{e9b}) is proportional to an
expression similar to (\ref{e11}) where $\beta_d$ is replaced by
$\beta_u$.  However, expression (\ref{e9b}) is here more appropriate
since its position-dependent version (\ref{e9}) is valid for all
$X_u<0$.  From (\ref{e9b}) one gets for real $Q_\ell$
\begin{equation}\label{e9c}
|{\cal U}_{\ell}|^2-|{\cal W}_{\ell}|^2
\underset{X_u \to -\infty}{\longrightarrow}
8\,\frac{{E}_{\ell}}{|{\cal D}_\ell|^2}
\left(\frac{\varepsilon_u}{Q_\ell}\right)^2,
\end{equation}
where ${E}_{\ell}=\varepsilon_u-\m_u Q_\ell$. In the following, the
constant ${\cal D}_\ell$ will be chosen such that $J_\ell=\pm 1$ for
real $Q_\ell$ (see the discussion at the end of Sec.
\ref{downstream}; one has here $J_{u|{\rm out}}=-1$ and $J_{u|{\rm
      in}}=+1$). Also, we chose ${\cal D}_{u|{\rm in}}$ on the positive
    imaginary axis in the complex plane and ${\cal D}_{u|{\rm out}}$
    on the negative imaginary axis. This implies that, for real
    $Q_\ell$,
\begin{equation}\label{normr}
{\cal D}_\ell = \sqrt{8} \, {\rm i} \, \frac{Q_\ell}{|Q_\ell|} \,
|{E}_{\ell}V_g(Q_\ell)|^{\frac{1}{2}}
\left|\frac{\varepsilon_u}{Q_\ell}\right|.
\end{equation}
The particular choice of phase in (\ref{normr}) is based on
aesthetic grounds: it ensures that the $\omega\to 0$ limit of the
eigen-function (\ref{e9b}) upstream the horizon in the waterfall and
delta peak configurations has the same phase as its equivalent for
the flat profile configuration. This will make it possible in Appendix
\ref{AppB} to obtain formulae valid for all three types of
configurations [Eqs. (\ref{b4}) and (\ref{b5})].

If $Q_\ell$ is complex, the expression is more complicated; we write
it here for completeness. One takes
\begin{align}\label{normq}
\notag {\cal D}_\ell
& = |V_g(Q_\ell)|^{\frac{1}{2}} \times \Big| 8 \, \mbox{Re} [
{E}_{\ell} ( \varepsilon_u/Q_\ell )^2 ] \\
\notag & + 4 \, \varepsilon_u
\cos^2\theta \, \frac{(Q_\ell-Q_\ell^*)^2}{|Q_\ell|^2} \\ 
& + 2 {\rm i} \, \varepsilon_u
\cos\theta \, \frac{Q_\ell-Q_\ell^*}{|Q_\ell|^2}
(Q_\ell^2+(Q_\ell^*)^2) \Big|^{\frac{1}{2}}.
\end{align}
This expression is clearly real and it ensures that, as in the
downstream region, the normalization (\ref{norme}) is fulfilled for
all $Q_\ell$.

\subsection{Matching at $x=0$}\label{matching}

Let us denote
\begin{equation}\label{e13}
\Xi_\alpha(x) = \begin{pmatrix}
\exp({\rm i} \m_\alpha X_\alpha) \bar{u}_\alpha(x) \\
\exp(-{\rm i} \m_\alpha X_\alpha) \bar{w}_\alpha(x)
\end{pmatrix}
=
\begin{pmatrix}
u_\alpha(x) \\
w_\alpha(x)
\end{pmatrix},
\end{equation}
and
\begin{equation}\label{e13a}
\Xi_\ell(x) =
\begin{pmatrix}
\exp({\rm i} (Q_\ell+\m_\alpha) X_\alpha) {\cal U}_\ell(x) \\
\exp({\rm i} (Q_\ell-\m_\alpha) X_\alpha) {\cal W}_\ell(x)
\end{pmatrix}.
\end{equation}
Remember that the index $\alpha$ is equal to either $u$ or $d$,
depending which side of the horizon one con\-si\-ders, whereas $\ell$
labels the eigen-modes of Eq. (\ref{e7}): $\ell\in\{u|{\rm in},
u|{\rm out}, u|{\rm eva}, d1|{\rm in}, d1|{\rm out}, d2|{\rm in},
d2|{\rm out}, d|{\rm eva} \}$. More precisely, $\Xi_u$ describes the
excitations in the sub\-so\-nic re\-gion; it is a linear
combination of $\Xi_{u|{\rm in}}$, $\Xi_{u|{\rm out}}$ and
$\Xi_{u|{\rm eva}}$. $\Xi_d$, which describes the same excitation in
the supersonic region, is a linear combination of $\Xi_{d1|{\rm
in}}$, $\Xi_{d1|{\rm out}}$, $\Xi_{d2|{\rm in}}$, $\Xi_{d2|{\rm
out}}$ and $\Xi_{d|{\rm eva}}$.

Then the matching conditions at the horizon read
\begin{equation}\label{e14}
\Xi_u(0) = \Xi_d(0),
\end{equation}
and
\begin{equation}\label{e15}
\frac{\hbar^2}{2m}\left[
\frac{{\rm d}\Xi_d}{{\rm d}x}(0) -
\frac{{\rm d}\Xi_u}{{\rm d}x}(0)
\right]
=
\Lambda\,\Xi_u(0).
\end{equation}
In the case of the flat profile or of the
waterfall configuration, Eq. (\ref{e15}) also holds,
but then $\Lambda=0$.

\subsection{The scattering modes}\label{scattering-modes}

Amongst all the possible modes described in Sec. \ref{matching} as
linear combinations of the $\Xi_\ell$'s, we are primarily interested
in the scattering modes. These are the three modes which are impinging on
the horizon along one of the three possible ingoing channels: $u|{\rm in}$,
$d1|{\rm in}$ or $d2|{\rm in}$. Each of these ingoing waves gives rise to
transmitted and reflected waves which, together with the initial
ingoing component, form what we denote as a ``scattering mode''. It
is natural to label these modes according to their incoming
channels, but since each mode includes
more than the ingoing wave that generates it, for avoiding confusion
in the notations, we use capital letters and denote the scattering
modes as $\Xi^{\sss U}$, $\Xi^{\sss D1}$ and $\Xi^{\sss D2}$.

\begin{figure}
\includegraphics*[width=0.99\linewidth]{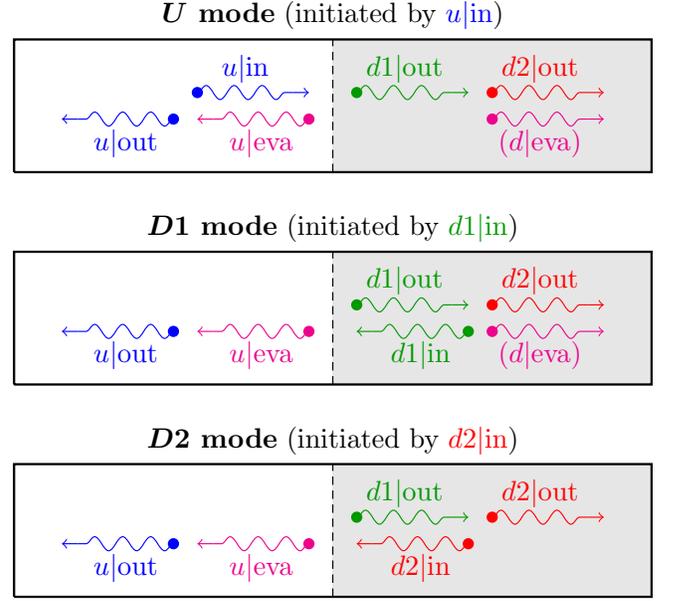}
\caption{(Color online) The scattering modes. The color code is the
  same as in Fig. \ref{fig3}. The additional purple wiggly lines
  correspond to evanescent channels. For $\omega>\Omega$, the $D2$
  mode disappears and the channel $d2|{\rm out}$ is replaced by
  $d|{\rm eva}$ in the two upper panels ($U$ and $D1$ modes). The
  region corresponding to the interior of the black hole is shaded as
  in the previous figures.}\label{fig5}
\end{figure}

For concreteness, we now give the expression of each of the scattering
modes. Each mode has a different analytical expression on each side
of the horizon. According to our conventions we denote these
expressions as $\Xi^{\sss U}_u(x)$, $\Xi^{\sss D1}_u(x)$ and
$\Xi^{\sss D2}_u(x)$ in the upstream region and $\Xi^{\sss U}_d(x)$,
$\Xi^{\sss D1}_d(x)$ and $\Xi^{\sss D2}_d(x)$ in the downstream
one. Specifically, one has
\begin{equation}\label{e17}
\begin{array}{l}
\vspace{2mm}
\Xi_{u}^{\sss U}=\Xi_{u|\mathrm{in}}+S_{u,u}\Xi_{u|\mathrm{out}}
+S_{u,u}^{\mathrm{eva}}\Xi_{u|\mathrm{eva}}, \\
\Xi_{d}^{\sss U}=S_{d1,u}\Xi_{d1|\mathrm{out}}+
\Theta(\Omega-\omega)S_{d2,u}\Xi_{d2|\mathrm{out}} \\
\vspace{2mm}
\hphantom{\textrm{}\Xi_{d}^{\sss U}=S_{d1,u}\Xi_{d1|\mathrm{out}}\textrm{}}+
\Theta(\omega-\Omega)S_{d,u}^{\mathrm{eva}}\Xi_{d|\mathrm{eva}}, \\
\vspace{2mm}
\Xi_{u}^{\sss D1}=S_{u,d1}\Xi_{u|\mathrm{out}}
+S_{u,d1}^{\mathrm{eva}}\Xi_{u|\mathrm{eva}}, \\
\Xi_{d}^{\sss D1}=\Xi_{d1|\mathrm{in}}+S_{d1,d1}\Xi_{d1|\mathrm{out}} \\
\hphantom{\textrm{}\Xi_{d}^{\sss D1}=\Xi_{d1|\mathrm{in}}\textrm{}}
+\Theta(\Omega-\omega)S_{d2,d1}\Xi_{d2|\mathrm{out}} \\
\vspace{2mm}
\hphantom{\textrm{}\Xi_{d}^{\sss D1}=\Xi_{d1|\mathrm{in}}\textrm{}}
+\Theta(\omega-\Omega)S_{d,d1}^{\mathrm{eva}}\Xi_{d|\mathrm{eva}}, \\
\vspace{2mm}
\Xi_{u}^{\sss D2}=\Theta(\Omega-\omega)(S_{u,d2}\Xi_{u|\mathrm{out}}
+S_{u,d2}^{\mathrm{eva}}\Xi_{u|\mathrm{eva}}), \\
\Xi_{d}^{\sss D2}=\Theta(\Omega-\omega)(\Xi_{d2|\mathrm{in}}
+S_{d1,d2}\Xi_{d1|\mathrm{out}} \\
\hphantom{\textrm{}\Xi_{d}^{\sss D2}=
\Theta(\Omega-\omega)(\Xi_{d2|\mathrm{in}}\textrm{}}
+S_{d2,d2}\Xi_{d2|\mathrm{out}}).
\end{array}
\end{equation}
For legibility, the $x$ dependence of the $\Xi$ vectors has not been
displayed in the equations. The three different modes are displayed in
a pictorial way in Fig. \ref{fig5} where the
purple wiggly lines correspond to the evanescent modes $u|{\rm eva}$ and
$d|{\rm eva}$ that cannot be
represented in Fig. \ref{fig3}. Note
that when $\omega>\Omega$, the outgoing $d2$ wave (involved in the $U$
and $D1$ modes) becomes evanescent and the $D2$ mode disappears,
because the incident seed for this mode (the propagating $d2|{\rm in}$
wave) disappears. This is taken care of in formulae (\ref{e17}) by the
Heaviside functions $\Theta(\Omega-\omega)$ and $\Theta(\omega-\Omega)$.

The $S$-coefficients ($S_{u,u}$, $S_{u,u}^{\rm eva}$, etc.) in
Eqs. (\ref{e17}) are complex and do not depend on $x$ (they do depend
on $\omega$ though). For each of the three scattering modes, one has
four such coefficients which, once the incident channel is fixed, are
determined by solving the $4 \times 4$ system of linear equations
(\ref{e14}) and (\ref{e15}): hence, the $S$-parameters depend on the
configuration considered (flat profile, delta peak or waterfall).
Physically, the square moduli $|S_{\nu',\nu}(\omega)|^2$ of the
$S$-matrix elements give the transmission or reflection coefficients for
a $\nu$-ingoing mode of energy $\hbar\omega$ which scatters into an
$\nu'$-outgoing mode at the same energy.

Current conservation can be written in a simple matrix form
provided the normalization of the real modes is defined in such a way
that $J_\ell=\pm 1$ (as done in Secs. \ref{downstream} and
\ref{upstream}). Defining, for $\omega<\Omega$, the $S$-matrix as
\begin{equation}\label{e18}
S=
\begin{pmatrix}
S_{u,u} & S_{u,d1} & S_{u,d2} \\
S_{d1,u} & S_{d1,d1} & S_{d1,d2} \\
S_{d2,u} & S_{d2,d1} & S_{d2,d2}
\end{pmatrix},
\end{equation}
current conservation reads
\begin{equation}\label{e19}
S^\dagger \eta S = \eta = S \eta S^\dagger, \quad \mbox{where} \quad
\eta = {\rm diag}(1,1,-1).
\end{equation}
The coefficients such as $S_{d,u}^{\rm eva}$ are not involved in
current conservation since the evanescent waves carry no current. Note
that for $\omega>\Omega$, the $S$-matrix is $2 \times 2$ because, the
outgoing $d2$ mode -- being evanescent in this case -- is not involved
in current conservation: one simply has the usual unitarity relation
$S^\dagger S = {\rm diag}(1,1)$. We have checked that our results for
the scattering matrix indeed fulfill the $\eta$-unitarity condition
(\ref{e19}) for $\omega<\Omega$ (and the unitarity condition for
$\omega>\Omega$).

In the following we will need to determine the low-$\omega$ behavior of the
components of the $S$-matrix. In the three configurations we
considered, we always find that, for $\nu=u, d1, d2$,
one has
\begin{align}\label{loww}
\notag&S_{\nu,u} = f_{\nu,u} +
h_{\nu,u} \varepsilon_u + {\cal O}(\varepsilon_u^2), \\
&S_{\nu,d1} = \frac{f_{\nu,d1}}{\sqrt{\varepsilon_u}} +
h_{\nu,d1}\sqrt{\varepsilon_u} + {\cal O}(\varepsilon_u^{3/2}) , \\
\notag&S_{\nu,d2} = \frac{f_{\nu,d2}}{\sqrt{\varepsilon_u}} +
h_{\nu,d2}\sqrt{\varepsilon_u} + {\cal O}(\varepsilon_u^{3/2}),
\end{align}
where $\varepsilon_u=\hbar\omega/(m c_u^2)$ and the $f$'s and the $h$'s
are dimensionless complex numbers. We determined them analytically in
the three configurations we considered. The relevant formulae are
given in Appendix \ref{AppA}.

\subsection{Quantization}\label{quantization}

The field operator $\hat{\psi}(x,t)$
associated in the Heisenberg representation
to the elementary excitations on top of the background [as
defined by Eq. (\ref{e5b})] is expanded over the scattering modes:
\begin{align}\label{q1}
\notag \hat{\psi}(x,t) &= 
\ep^{{\rm i} k_{\alpha} x} \int_{0}^{\infty} \frac{{\rm d}\omega}{\sqrt{2\pi}}
\sum_{\sss L\in\{U,D1\}} \Big[ \bar{u}_{\sss L}(x,\omega) 
\ep^{-{\rm i} \omega t} \hat{a}_{\sss L}(\omega) \\
\notag &\hphantom{= \ep^{{\rm i} k_{\alpha} x} \int_{0}^{\infty} 
\frac{{\rm d}\omega}{\sqrt{2\pi}} \sum_{\sss L\in\{U,D1\}}}
+ \bar{w}_{\sss L}^{*}(x,\omega) \ep^{{\rm i} \omega t} 
\hat{a}_{\sss L}^{\dag}(\omega) \Big] \\
\notag &+ \ep^{{\rm i} k_{\alpha} x} \int_{0}^{\Omega} 
\frac{{\rm d}\omega}{\sqrt{2\pi}}
\Big[ \bar{u}_{\sss D2}(x,\omega) \ep^{-{\rm i} \omega t} 
\hat{a}_{\sss D2}^{\dag}(\omega) \\
&\hphantom{= \ep^{{\rm i} k_{\alpha} x} \int_{0}^{\Omega} 
\frac{{\rm d}\omega}{\sqrt{2\pi}}}
+ \bar{w}_{\sss D2}^{*}(x,\omega) \ep^{{\rm i} \omega t} 
\hat{a}_{\sss D2}(\omega) \Big],
\end{align}
where we have written explicitly the $\omega$ dependence. The
$\hat{a}_{\sss L}^\dagger(\omega)$'s create an excitation of energy
$\hbar\omega$ in one of the three scattering modes ($U$, $D1$ or
$D2$).  They obey the following commutation relation:
\begin{equation}\label{q2}
[\hat{a}_{\sss L}(\omega),\hat{a}^\dagger_{\sss L'}(\omega')] =
\delta_{\sss L,L'}\delta(\omega-\omega').
\end{equation}
From expression (\ref{q1}) one sees that the $D2$ mode (which
originates from the negative norm $d2|{\rm in}$ channel) is quantized
in a non standard way: the role of the creation and annihilation
operators is exchanged compared to the $U$ and $D1$ modes. Using the
current conservation relation (\ref{e19}), one can show that this
choice of quantization is necessary for fulfilling the appropriate
Bose commutation relation of the $\hat\psi$ operator:
\begin{equation}\label{q3}
[\hat{\psi}(x,t),\hat{\psi}^\dagger(x',t)]=\delta(x-x').
\end{equation}

\section{Radiation spectrum}\label{radiation}

The Hawking signal corresponds to emission of radiation from the
interior toward the exterior of the black hole \cite{Hawking}. In our
specific case the energy current associated to emission of elementary
excitations is (cf. \cite{Kag03})
\begin{equation}\label{rs3}
\hat{\Pi}(x,t) = -\frac{\hbar^2}{2m} \,
\partial_t\hat\Phi^\dagger(x,t) \, \partial_x\hat\Phi(x,t) + {\rm h.c.},
\end{equation}
where ``h.c.'' stands for ``hermitian conjugate''.
From expressions (\ref{e5b}) and (\ref{q1})
one can write the average current $\Pi(x) = \langle \hat{\Pi}(x,t) \rangle$
under the form
\begin{equation}\label{rs4}
\Pi(x) = \int_0^\infty \frac{{\rm d}\omega}{2\pi} \,
\hbar\omega \, J(x,\omega),
\end{equation}
where $J(x,\omega)$ [and accordingly $\Pi(x)$]
can be separated in a zero temperature part
$J_0(x,\omega)$ [$\Pi_0(x)$]
and a ``thermal part'' $J_{\sss T}(x,\omega)$ [$\Pi_{\sss T}(x)$] with
\begin{align}\label{rs5}
\notag J_0(x,\omega) = \frac{\hbar}{2m} \Big[
&-\sum_{\sss L\in\{U,D1\}}
\bar{w}^*_{\sss L} (k_\alpha + {\rm i} \partial_x) \bar{w}_{\sss L} \\
\notag &+ \Theta(\Omega-\omega) \bar{u}^*_{\sss D2} 
(k_\alpha - {\rm i} \partial_x) \bar{u}_{\sss D2}
\Big] \\ &+ {\rm c.c.}
\end{align}
and
\begin{equation}\label{rs6}
J_{\sss T}(x,\omega)
=
\sum_{\sss L\in\{U,D1,D2\}}
J_{\sss L}(x,\omega) \, n_{\sss L}(\omega),
\end{equation}
where
\begin{equation}\label{rs7}
J_{\sss L} =
\frac{\hbar}{2m}
\Big[ \bar{u}^*_{\sss L} (k_\alpha - {\rm i} \partial_x) \bar{u}_{\sss L}
- \bar{w}^*_{\sss L} (k_\alpha + {\rm i} \partial_x) \bar{w}_{\sss L} \Big]
+ {\rm c.c.}
\end{equation}
and $n_{\sss L}(\omega) = \langle \hat{a}^\dagger_{\sss L}(\omega)
\hat{a}_{\sss L}(\omega) \rangle$ is the occupation number of the mode
$L$.  Note that in expression (\ref{rs6}) the $D2$ mode contributes
only for $\omega<\Omega$.  Comparing the expression (\ref{rs7}) with
(\ref{cc4}) one sees that $J_{\sss L}$ is the conserved current
carried by a scattering mode $L$; it is $x$ independent for the
stationary flows we consider in the present work.

Note for avoiding confusion that what we call a ``zero temperature
term'' is the contribution to the Hawking signal that exists even when
the system is at zero temperature. It will be described below
(Sec.~\ref{hawking-temp}) by an effective radiation
temperature $T_{\rm\sss H}$ (the Hawking tem\-pe\-ra\-tu\-re)
which is not the temperature of the BEC.

\subsection{Energy current in a black hole configuration}

For large and negative $x$ (i.e., deep in the subsonic region)
we show in Appendix \ref{AppC} 
that formulae (\ref{e17}) and (\ref{rs5}) yield the very natural result
\begin{equation}\label{rs13}
\Pi_0 = - \int_0^\Omega \frac{{\rm d}\omega}{2\pi} \, \hbar\omega \,
|S_{u,d2}(\omega)|^2.
\end{equation}
The minus sign in this equation indicates that the e\-ner\-gy current is
directed toward $-\infty$, as clearly seen from Appendix
\ref{AppC}. If one computes the energy current for a point deep in the
supersonic region (i.e., for $x$ large and positive) one gets the same
result as (\ref{rs13}), in agreement with the con\-ser\-va\-tion of the
energy flux in a stationary con\-fi\-gu\-ra\-tion. Note that the zero
temperature radiation $\Pi_0$ va\-ni\-shes in absence of black hole, as
expected. In presence of a black hole, the integral gives a finite
result, corres\-pon\-ding to a Hawking signal emitted even for $T=0$. This
remark, together with the specific form of Eq. (\ref{rs13}), shows
that one needs two ingredients for having a $T=0$ Hawking radiation
from a black hole: (i) a $d2|{\rm in}$ mode and (ii) a $d2
\leftrightarrow u$ mode conversion, i.e., a non zero $S_{u,d2}$
coefficient. Remember that condition (i) is fulfilled only because the
dispersion relation in a supersonic flow bends down at high $q$ (see
Fig. \ref{fig3}, bottom panel), which is a consequence of the non
linear behavior of the Bogoliubov dispersion relation (\ref{bog}). As
discussed in Sec. \ref{general}, the $d2$ incoming channel corresponds
to waves whose group velocity in the frame of the condensate is larger
than $V_d$. It is thus not surprising that these ``fast'' modes are
involved in the Hawking radiation since they are able to overrun the
flow of velocity $V_d$ and thus to escape the black hole. This is
clearly a flaw of the BEC analogy of gravitational black holes: point
(i) is certainly not fulfilled in the gravitational case since the
group velocity of photons is a constant (the speed of light). As a
result the number of ingoing and outgoing channels are not equal for
gravitational black holes and one cannot get a stationary description
of the Hawking effect: one has to take into account the dynamics of
the formation of the horizon.

From Eqs. (\ref{e17}) and (\ref{rs6}), the term $J_{\sss T}(x,\omega)$
can be rewritten as
\begin{align}\label{rs15}
\notag J_{\sss T}(x,\omega)
&= \frac{2}{c_u \xi_u} \Big[
n_{\sss U}(\omega) (1-|S_{u,u}|^2) \\
\notag &- n_{\sss D1}(\omega) |S_{u,d1}|^2 -
n_{\sss D2}(\omega) |S_{u,d2}|^2 \Big] \\
\notag &= \frac{2}{c_u \xi_u} \Big\{
[n_{\sss U}(\omega) - n_{\sss D1}(\omega)] |S_{u,d1}|^2 \\
&- [n_{\sss U}(\omega) + n_{\sss D2}(\omega)] |S_{u,d2}|^2 \Big\}.
\end{align}
In absence of black hole, the $S_{u,d2}$ term in (\ref{rs15})
disappears. As a result, at thermal equilibrium,
i.e., when $n_{\sss U}(\omega) = n_{\sss D1}(\omega)$ is a thermal
Bose occupation number of the form
\begin{equation}\label{bosefactor}
n_{\sss T}(\omega) = 
\frac{1}{\exp\big(\frac{\hbar\omega}{k_{\rm\sss B}T}\big)-1},
\end{equation} 
one has $J_{\sss T}(x,\omega) = 0$ [this is most easily seen from the
last expression of $J_{\sss T}$ in Eq. (\ref{rs15})]. This is a very
pleasant result demonstrating that at thermal equilibrium there is no
Hawking radiation at all for any type of configuration connecting two
asymptotically subsonic regions.

In the case where an acoustic horizon is present, a finite temperature
configuration may be reached in the manner presented in
Ref. \cite{correlations}: one branches the black hole configuration
adiabatically starting from a system initially uniformly subsonic at
thermal equilibrium. This changes the dispersion relation in the
supersonic part, but not the occupation number of the adiabatically
modified modes (see the discussion in \cite{Rec09}). In this case
(\ref{rs15}) yields a finite value for the $\Pi_{\sss T}$ term
(because $S_{u,u}$, $n_{\sss D1}$ and $n_{\sss D2}$ are regular at low
$\omega$).

\subsection{Hawking temperature}\label{hawking-temp}

The zero temperature radiation $\Pi_0$ as given by (\ref{rs13})
corresponds to an emission spectrum given by
$|S_{u,d2}(\omega)|^2$. Can this be described by an effective
tem\-pe\-ra\-tu\-re, i.e., can $|S_{u,d2}(\omega)|^2$ be approximated
by a factor of the type $\Gamma \times n_{\sss T_{\rm\sss
    H}}(\omega)$, where $T_{\rm\sss H}$ is the effective
tem\-pe\-ra\-tu\-re of radiation? We address this question in the
present subsection.

\begin{figure}
\includegraphics*[width=0.99\linewidth]{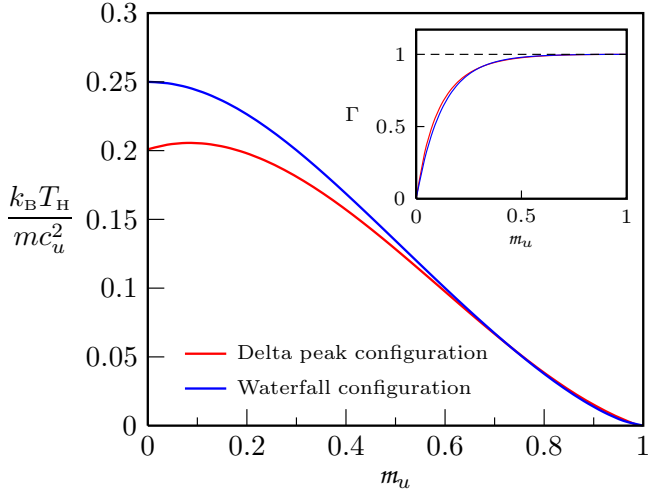}
\caption{Normalized Hawking temperature $k_{\rm\sss
B}T_{\rm{\sss H}}/(m c_u^2)$ as a function of the upstream Mach
number $\m_u=V_u/c_u$ for the delta peak and waterfall
configurations. The inset displays the grayness factor $\Gamma$ as
a function of $\m_u$ for these two con\-fi\-gu\-ra\-tions.}\label{figth}
\end{figure}

One could first argue that the addition of the ``grayness factor''
$\Gamma$ is an unnecessary complication of the fit of $|S_{u,d2}|^2$
by a thermal spectrum. Indeed, in most cases, $\Gamma$ is found to be
close to 1, but this is not generally true (see the discussion below
and the inset of Fig. \ref{figth}) and this is the reason why we keep
a certain degree of ``grayness'' in the present analysis.

Obviously, the identification of $|S_{u,d2}|^2$ with a Bose thermal
factor can only be approximate because, whereas a term such as
$n_{\sss T_{\rm\sss H}}(\omega)$ is finite for all
$\omega\in\mathbb{R}^+$, $|S_{u,d2}|^2$ abruptly cancels for
$\omega>\Omega$. Nonetheless one can try to find the best possible
approximation by comparing the low-$\omega$ expansion of
$|S_{u,d2}|^2$ [from Eqs. (\ref{loww})] with $\Gamma \times n_{\sss
  T_{\rm\sss H}}(\omega) = \Gamma[k_{\rm\sss B}T_{\rm\sss
  H}/(\hbar\omega) - \frac{1}{2} + {\cal O}(\omega)]$. This
immediately yields
\begin{equation}\label{rs18}
\Gamma = -4\,
{\rm Re}(f_{u,d2}^*h_{u,d2}) \quad \mbox{and} \quad
\frac{k_{\rm\sss B}T_{\rm\sss H}}{m c_u^2} =
\frac{|f_{u,d2}|^2}{\Gamma}.
\end{equation}
The analytical expressions obtained in Appendix \ref{AppA} for
$f_{u,d2}$ and $h_{u,d2}$ in the different configurations we consider
yield the following estimates of the Hawking temperature:
\begin{equation}\label{rs19}
\frac{k_{\rm\sss B}T_{\rm\sss H}}{m c_u^2} = \left\{
\begin{array}{ll}
\vspace{2mm}
\displaystyle{\frac{1}{2} \frac{\m_u^2}{\m_d}
\frac{(1-\m_u^2)(\m_d^2-1)^{\frac{3}{2}}}{\m_d^2-\m_u^2}}
& \mbox{(flat profile)}, \\
\displaystyle{\frac{1}{2}
\frac{(1-\m_u^4)^{\frac{3}{2}}}
{(2+\m_u^2)(1+2\m_u^2)}}
& \mbox{(waterfall)}.
\end{array}\right.
\end{equation}
We do not display here the formula for the delta peak configuration
because it is too cumbersome. Instead we show in Fig. \ref{figth} the
corresponding curve relating $T_{\rm\sss H}$ to $\m_u$ in the delta
peak configuration and compare it with the results of the waterfall
configuration. One first notices from the figure that $T_{\rm\sss
  H}\to 0$ when $\m_u\to 1$: this is expected because in this case
the horizon disappears. One sees also that $T_{\rm\sss H}$ remains
finite in the limit $\m_u\to 0$ in both configurations. However, this
limit is singular in the sense that it corresponds to a very peculiar
flow. For instance, in the waterfall configuration, the analysis of
Sec. \ref{waterfall} shows that the flow with $\m_u=0$ is observed for
a step with $U_0\to\infty$, and has downstream a zero density and an
infinite velocity: the corresponding flow pattern is most probably
unreachable. Moreover, one sees from the inset of Fig. \ref{figth}
(and also from the analytical expression given in Appendix \ref{AppA})
that in this case $\Gamma\to 0$ and the expected signal
disappears. In the remainder of this work we rather consider a typical
setting with $\m_u=0.5$. For this value of $\m_u$ one gets $\m_d\simeq
1.83$ and $\Gamma\simeq 0.977$ in the delta peak configuration,
$\m_d=4$ and $\Gamma\simeq 0.980$ in the waterfall configuration.

\begin{figure}
\includegraphics*[width=0.99\linewidth]{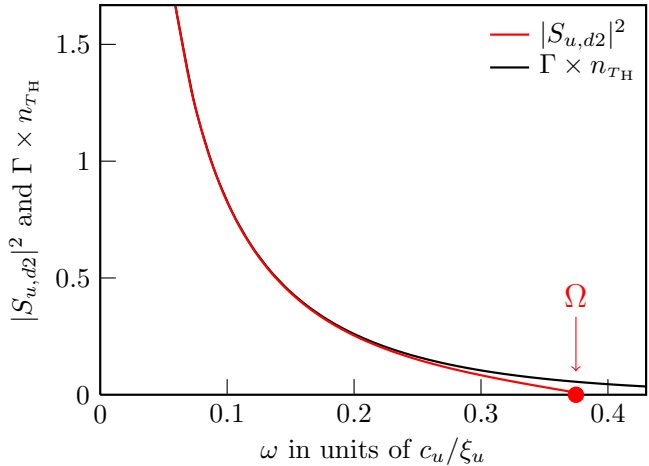}
\caption{(Color online) Radiation spectrum in the delta peak
configuration. Red curve: $|S_{u,d2}|^2$ as a function of the
dimensionless quantity $\omega\xi_u/c_u$. Black curve:
$\Gamma \times n_{\sss T_{\rm\sss H}}(\omega)$. The plot is drawn for
$\m_u=0.5$; in this case $k_{\rm\sss B}T_{\rm\sss H}
/(m c_u^2)\simeq 0.128$ and $\Gamma\simeq 0.977$. The difference between
$|S_{u,d2}|^2$ and $\Gamma \times n_{\sss T_{\rm\sss H}}$ is maximum when
$\omega=\Omega$ and is close to 0.06 in this case (for the chosen
value of $\m_u$ one has $\Omega\xi_u/c_u\simeq 0.369$).}\label{fig11}
\end{figure}

Once $T_{\rm\sss H}$ is determined through the above low energy
analysis, one should check, as done, e.g., in Ref. \cite{Mac09}, if
the approximation of $|S_{u,d2}|^2$ with a thermal spectrum is
accurate in the whole emission window $\omega \in [0,\Omega]$. This is
done in Fig. \ref{fig11} in the case of the delta peak
configuration. One sees from the figure that the overall agreement is
quite good. The same good agreement is obtained for the waterfall and
flat profile configurations, and this legitimates the definition of a
Hawking temperature in the three configurations considered in the
present work. This was not \textit{a priori} obvious because the
concept of Hawking temperature is of semi-classical origin (see, e.g.,
\cite{Vol03}) and the three configurations we consider being
discontinuous, one could fear that a semi-classical analysis would
fail. We see the relevance of the concept of Hawking temperature as a
confirmation that the configurations considered here are typical for
observing the Hawking effect. In the next section we will draw the
same conclusion from a study of the density correlations.

From (\ref{rs19}) and Fig. \ref{figth} one gets an order of magnitude
$k_{\rm\sss B}T_{\rm\sss H} / (m c^2_u) \sim 0.1$, i.e., typically
$T_{\rm\sss H} \sim 10$ nK.  Since the temperature in typical
experiments is rather of the order of the chemical potential $mc^2_u$
(i.e., around 100 nK), the Hawking radiation will be lost in the thermal
noise and very difficult to identify. This is the reason why density
correlations have been proposed in Ref. \cite{correlations} as a tool
for identifying the Hawking effect. We thus consider two-body
correlations in the next section.

\section{Correlations}\label{correl}

The connected two-body density matrix is defined by \cite{qu-optics}
\begin{align}\label{q4}
\notag g^{(2)}&(x_1,x_2) \\ \notag &=
\langle \hat{\Phi}^\dagger(x_1,t) \hat{\Phi}^\dagger(x_2,t)
\hat{\Phi}(x_1,t) \hat{\Phi}(x_2,t) \rangle \\
& - \langle \hat{\Phi}^\dagger(x_1,t) \hat{\Phi}(x_1,t) \rangle
\langle \hat{\Phi}^\dagger(x_2,t) \hat{\Phi}(x_2,t) \rangle.
\end{align}
$g^{(2)}$ is time independent because we work in a stationary
configuration. In (\ref{q4}) the average is taken either on the
ground state or over a statistical ensemble. $g^{(2)}$ is directly
related to the density correlations in the system, this can be seen
by rewriting Eq. (\ref{q4}) under the form
\begin{align}\label{pe2}
\notag g^{(2)}(x_1,x_2) &= \langle \hat{n}(x_1,t) \hat{n}(x_2,t) \rangle
- \langle \hat{n}(x_1) \rangle \langle \hat{n}(x_2) \rangle \\
& - \delta(x_1-x_2) \langle \hat{n}(x_1) \rangle.
\end{align}
The last term in the right-hand side (r.h.s.) of (\ref{pe2}) is the
Poissonian fluctuation term originating from the discreteness of the
particles \cite{Landau5}. Written under this form, $g^{(2)}$ is
sometimes denoted as the cluster function.

For a system at thermal equilibrium in the grand canonical ensemble,
one has
\begin{align}\label{corr1}
\notag n(x) &= \langle \hat{n}(x) \rangle \\ &
= \frac{1}{{\cal Z}} \, \mbox{Tr} \left\{ \hat{n}(x)
\exp\left[-\frac{1}{k_{\rm\sss B}T} (\hat{H} - \mu \hat{N})\right] \right\},
\end{align}
where ${\cal Z}= \mbox{Tr}\{\exp[-(\hat{H}-\mu\hat{N})/(k_{\rm\sss
  B}T)]\}$ is the partition function. Deriving expression
(\ref{corr1}) with respect to $\mu$, one gets
\begin{equation}\label{corr2}
k_{\rm\sss B}T \frac{\partial n(x)}{\partial\mu} =
\langle \hat{n}(x) \hat{N} \rangle -
\langle \hat{n}(x)\rangle \langle \hat{N} \rangle.
\end{equation}
Since $\hat{N} = \int_{\mathbb{R}} {\rm d}x' \, \hat{n}(x')$, property
(\ref{corr2}) and expression (\ref{pe2}) yield the following sum rule:
\begin{equation}\label{corr3}
\int_{\mathbb{R}} {\rm d}x' \, g^{(2)}(x,x') = -n(x) + k_{\rm\sss B}T
\frac{\partial n(x)}{\partial\mu}.
\end{equation}
For a homogeneous system, this sum rule is a standard thermodynamic
result \cite{Landau5} which can be shown to be equivalent to the
compressibility sum rule (whose de\-fi\-ni\-tion is given for instance in
Ref. \cite{PinesNozieres}). Formula (\ref{corr3}) is a generalization
to inhomogeneous systems; it has been used in Ref. \cite{Arm11} for
witnessing quasi-condensation through a study of density fluctuations
and in Ref. \cite{Zho11} to propose an universal thermometry for
quantum simulations.

In the remainder of this section we concentrate on the $T=0$ case,
postponing the discussion of finite temperature to a future
publication. Fulfillment of the sum rule (\ref{corr3}) is a strong test
of the validity of the Bogoliubov approach used in the present
work. We give in Sec. \ref{g2-bh} the leading order
contributions to $g^{(2)}$ and explain in Appendix \ref{AppB} how we
use them in order to check that the $T=0$ version of the sum rule
(\ref{corr3}) is indeed verified for $|x|\to\infty$ (i.e., far from
the horizon).

From the Bogoliubov expansion (\ref{e5b}), one gets at leading order
\begin{align}\label{q4a}
\notag g^{(2)}(x_1,x_2) & =
\Phi(x_1) \Phi^*(x_2)
\langle \hat{\psi}^\dagger(x_1,t) \hat{\psi}(x_2,t) \rangle \\
\notag & + \Phi(x_1) \Phi(x_2)
\langle \hat{\psi}^\dagger(x_1,t) \hat{\psi}^\dagger(x_2,t) \rangle \\
& + \rm{c.c.}.
\end{align}
For $i=1$ or $2$, we write $\Phi(x_i)=\sqrt{n_i}\exp({\rm i}k_i
x_i)\phi_i(x_i)$, where $n_i=n_u$ ($n_d$), $k_i=k_u$ ($k_d$) and
$\phi_i=\phi_u$ ($\phi_d$) if $x_i<0$ ($x_i>0$). We recall that $\phi_d$ is
defined in Eq. (\ref{down}) and $\phi_u$ is either equal to unity
(flat profile configuration) or given by Eq. (\ref{up}) (delta peak
and waterfall configurations). Based on the decomposition (\ref{q1}),
one can show (see Ref. \cite{Rec09}) that Eq. (\ref{q4a}) yields
\begin{equation}\label{q5}
g^{(2)}(x_1,x_2) = \sqrt{n_1 n_2} \int_0^\infty \frac{{\rm d}\omega}{2\pi}
\, \gamma(x_1,x_2,\omega),
\end{equation}
where $\gamma(x_1,x_2,\omega)$ [and accordingly $g^{(2)}(x_1,x_2)$]
is conveniently separated in a zero temperature term $\gamma_0$ and a
remaining term $\gamma_{\sss T}$:
\begin{equation}\label{q5a}
\gamma(x_1,x_2,\omega)
= \gamma_0(x_1,x_2,\omega) + \gamma_{\sss T}(x_1,x_2,\omega).
\end{equation}
$g^{(2)}_0$ and $\gamma_0$ are the contributions evaluated from
(\ref{q4a}) and (\ref{q1}) which remain finite even in the $T=0$ case
where $n_{\sss L}(\omega) = \langle \hat{a}^\dagger_{\sss L}(\omega)
\hat{a}_{\sss L}(\omega) \rangle = 0$.  One has
\begin{align}\label{q6}
\notag \gamma_0(x_1,x_2,\omega) &=
\sum_{\sss L\in\{U,D1\}}
\tilde{w}^*_{\sss L}(x_1) \tilde{r}_{\sss L}(x_2) \\
& + \Theta(\Omega-\omega)
\tilde{u}^*_{\sss D2}(x_1) \tilde{r}_{\sss D2}(x_2)
+ {\rm c.c.},
\end{align}
with
\begin{equation}\label{q7}
\tilde{u}_{\sss L}(x_i) =
\phi_i^*(x_i) \bar{u}_{\sss L}(x_i), \quad
\tilde{w}_{\sss L}(x_i) =
\phi_i(x_i) \bar{w}_{\sss L}(x_i),
\end{equation}
and
\begin{equation}
\tilde{r}_{\sss L}(x_i) =
\tilde{u}_{\sss L}(x_i) + \tilde{w}_{\sss L}(x_i).
\end{equation}
The other contribution to (\ref{q5a}) is
\begin{equation}\label{q7a}
\gamma_{\sss T}(x_1,x_2,\omega) =  \!\!\!\! \sum_{\sss L\in\{U,D1,D2\}} \!\!\!\!
\tilde{r}_{\sss L}^*(x_1) \tilde{r}_{\sss L}(x_2)
n_{\sss L}(\omega) + {\rm c.c.},
\end{equation}
where it should be understood that the $D2$ contribution is only
present for $\omega<\Omega$.

We often display below the results not
for $g^{(2)}$ but for the dimensionless quantity $G^{(2)}$ defined as
\begin{equation}\label{q4b}
G^{(2)}(x_1,x_2) = \frac{g^{(2)}(x_1,x_2)}{n(x_1) n(x_2)}.
\end{equation}
Also, we will compute the correlations when $x_1$ and $x_2$ are both
far from the horizon. In this case, $\phi_i(x_i) = \exp({\rm i}\beta_i)$
and $G^{(2)}(x_1,x_2) = g^{(2)}(x_1,x_2)/(n_1 n_2)$.

\subsection{No black hole}

\begin{figure}
\includegraphics*[width=0.99\linewidth]{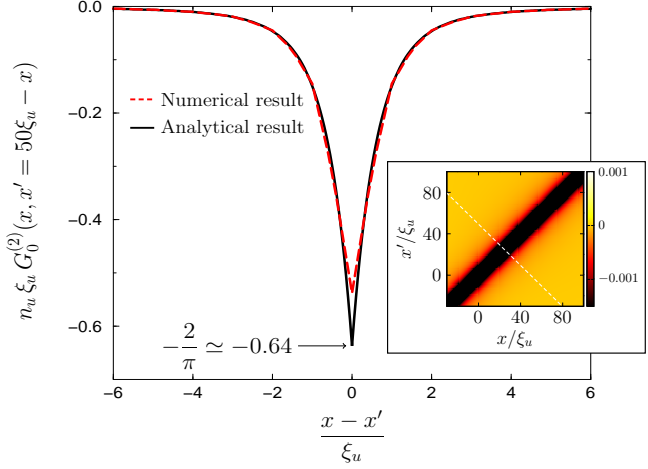}
\caption{(Color online) $\xi_u n_u G^{(2)}_0(x,x')$ computed
  analytically from Eq. (\ref{co2b}) (black solid line) compared with
  the numerics from the truncated Wigner method (red dashed line). The
  small discrepancy near $x-x'=0$ is due to numerical uncertainty and to
  the plotting procedure which introduces a small amount of smoothing
  of the raw numerical data.  The inset represents a color plot of the
  numerical results. The white dashed line is the line $x'=50\xi_u-x$
  along which we compare numerical and analytical results in the main
  plot.}\label{fig7}
\end{figure}

Before embarking in a general determination of $g^{(2)}$ for a black
hole configuration, we recall here the result for a uniform
fluid (density $n_u$) moving at constant subsonic velocity.
One gets from the no black hole version of (\ref{q6})
\cite{rem1}
\begin{align}\label{co1}
\notag \gamma_0&(x,x',\omega) \\
&= \sum_{\ell\in\{u|{\rm in},u|{\rm out}\}}
\tilde{w}_\ell^*(x) \tilde{r}_\ell(x') + {\rm c.c.} \\
&=
\sum_{\ell\in\{u|{\rm in},u|{\rm out}\}}
\frac{\frac{Q_\ell^2}{2|{E}_{\ell}|}-1}{2|V_g(Q_\ell)|} \,
\ep^{{\rm i} Q_\ell (X'_u-X_u)} + \rm{c.c.}.
\end{align}
This yields
\begin{equation}\label{co2b}
G^{(2)}_0(x,x') = \frac{1}{n_u \xi_u}
F\left(\frac{x-x'}{\xi_u}\right),
\end{equation}
where
\begin{equation}
F(z) = - \frac{1}{\pi z} \int_0^\infty {\rm d}t \, \frac{\sin(2 \, t \, z)}
{(1+t^2)^{3/2}}.
\end{equation}
This is the expected correlation in a quasi 1D condensate (see, e.g.,
Ref. \cite{Deu09} and references therein): $n_u \xi_u G^{(2)}_0(x,x')$
is a universal function of $z=(x-x')/\xi_u$. In particular, $n_u \xi_u
G^{(2)}_0(x,x)=F(0)=-2/\pi$ \cite{Gan03}. We evaluated the
fluctuations around the uniform profile numerically by means of the
truncated Wigner method for the Bose field \cite{Wal94,Ste98} already
used in \cite{correlations} for studying the same observable in a
black hole configuration. In Fig. \ref{fig7} we compare the analytical
form (\ref{co2b}) with the results of the numerical computation along
the cut displayed in the inset.  The excellent agreement is a good
test of the accuracy of the numerical method used in
Ref. \cite{correlations}.

We finally note here that $\int_{\mathbb{R}}{\rm d}z\,F(z)=-1$. This yields
\begin{equation}\label{co2d}
\int_{\mathbb{R}} {\rm d}x' \, g^{(2)}_0(x,x') = -n_u,
\end{equation}
which is a mere verification of the $T=0$ version of the sum rule
(\ref{corr3}) in this simple uniform setting.

The main correlation signal in the black hole configurations to be
studied soon is similar to the short range anti-bunching displayed in
Fig. \ref{fig7}. However, we will see that (i) its
precise shape is affected in presence of an acoustic horizon and
moreover, (ii) new long range correlations appear, which can be
interpreted as emission of correlated phonons \cite{correlations}.

\subsection{General formulae in presence of a black hole}\label{g2-bh}

We now turn to the study of zero temperature density fluctuations
around a sonic horizon. For simplicity, we only consider the case
where $x$ and $x'$ are far from the horizon: this makes it possible (i) to drop
the evanescent contributions to (\ref{q6}) and (ii) to avoid treating
the position dependence of the background density in the delta peak
and waterfall configurations. Part of these results were already
obtained in Ref. \cite{Rec09} (the ones valid when $x$ and $x'$ are
far one from the other) and here we gene\-ra\-li\-ze and correct some
misprints \cite{misprints}. We only display the most important
contributions to $\gamma_0$ which are both the larger ones and the
ones useful for fulfilling the $T=0$ version of the sum rule
(\ref{corr3}) when $x$ is far from the horizon [this version is
written explicitly in Eq. (\ref{b1bis})]. We note here that a similar
approach has previously been used in Ref. \cite{Bou03} for studying
phase fluctuations in a similar setting, with a description of the
scattering less elaborate than the one presented in
Sec. \ref{scattering-modes}.

Finally, we introduce the notations $\tilde{\cal U}_\ell$ and
$\tilde{\cal W}_\ell$ defined by
\begin{equation}\label{cbh1}
\begin{pmatrix}
\tilde{u}_\ell(x) \\ \tilde{w}_\ell(x) \end{pmatrix} =
\ep^{{\rm i} q_\ell x}
\begin{pmatrix}
\tilde{\cal U}_\ell(x) \\ \tilde{\cal W}_\ell(x)
\end{pmatrix},
\end{equation}
where $\tilde{u}_\ell(x)$ and $\tilde{w}_\ell(x)$ are defined in
Eqs. (\ref{q7}). We also introduce $\tilde{\cal R}_\ell =
\tilde{\cal U}_\ell + \tilde{\cal W}_\ell$.

\subsubsection{Case $x$ and $x' \to -\infty$}\label{outout}

We first consider the case where $x$ and $x'$ are both deep in the
subsonic region, i.e., outside the black hole and far from the
acoustic horizon.  From Eq. (\ref{q6}), one gets in this case
\begin{align}\label{c05}
\notag \gamma_0(x,x',\omega) & =
\tilde{{\cal W}}^*_{u|{\rm in}} \tilde{{\cal R}}_{u|{\rm in}}
\ep^{{\rm i} q_{u|{\rm in}} (x'-x)} \\
\notag & +
\tilde{{\cal W}}^*_{u|{\rm out}} \tilde{{\cal R}}_{u|{\rm out}}
\ep^{{\rm i} q_{u|{\rm out}} (x'-x)} \\
\notag & +
\Theta(\Omega-\omega)
|S_{u,d2}|^2
|\tilde{{\cal R}}_{u|{\rm out}}|^2
\ep^{{\rm i} q_{u|{\rm out}} (x'-x)} \\
& + \rm{c.c.}.
\end{align}
The contribution of the $S_{u,d2}$ term disappears when
$\omega>\Omega$ and this is the reason for the Heaviside factor
$\Theta(\Omega-\omega)$ in (\ref{c05}). If it were not for the
$S_{u,d2}$ term, (\ref{c05}) would be exactly equal to (\ref{co1}),
one would recover the same correlation as (\ref{co2b}) obtained in
absence of black hole and the contribution of (\ref{c05}) alone would
be enough to verify the sum rule (\ref{co2d}). Now the $S_{u,d2}$
term is not zero, and this means that the correlations in the vicinity of
the diagonal $x=x'$ are modified by the existence of the black
hole. This is similar to the results obtained by
Kravtsov and coworkers for non
standard ensembles of random matrices \cite{Fra09,Can95}. Indeed, for
fixed $x$, (\ref{c05}) alone is not able to fulfill the sum rule. The
addition of the non local correlations (\ref{c04c}) induced by the
Hawking emission will be necessary to this end, as advocated in
Ref. \cite{Fra09}.

\subsubsection{Cases ($x\to-\infty$ and $x'\to+\infty$) or
($x\to+\infty$ and $x'\to-\infty$)}\label{inout}

In the case where $x$ is deep in the upstream
region and $x'$ deep in the downstream one, we get
\begin{align}\label{c04c}
\notag \gamma_0(x,x',\omega) 
& = \Theta(\Omega-\omega) S^*_{u,d2} S_{d1,d2} \times \\
\notag & \hphantom{\textrm{}+\textrm{}} \tilde{\cal R}^*_{u|\rm{out}}
\tilde{\cal R}_{d1|\rm{out}}
\ep^{{\rm i} (q_{d1|\rm{out}}x' - q_{u|\rm{out}}x)} \\
\notag & + \Theta(\Omega-\omega) S^*_{u,d2} S_{d2,d2} \times \\
\notag & \hphantom{\textrm{}+\textrm{}} \tilde{\cal R}^*_{u|\rm{out}} 
\tilde{\cal R}_{d2|\rm{out}}
\ep^{{\rm i} (q_{d2|\rm{out}}x' - q_{u|\rm{out}}x)} \\
& + {\rm c.c.}.
\end{align}
If instead $x$ is deep in
the downstream supersonic region and $x'$ deep in the upstream
subsonic region, it suffices to exchange the roles of $x$ and $x'$ in
the above formula.

\subsubsection{Case $x$ and $x'\to+\infty$}\label{inin}

This is the case where $x$ and $x'$
are both deep in the downstream region (i.e., deep inside the black
hole). The leading order contribution to $g^{(2)}_0$ can be
separated in a diagonal part which depends only on $x-x'$ and a non
diagonal part. The diagonal part reads
\begin{align}\label{c06}
\notag \gamma_{0}^{\rm diag}(x,x',\omega) & =
\tilde{{\cal W}}^*_{d1|{\rm in}} \tilde{{\cal R}}_{d1|{\rm in}}
\ep^{{\rm i} q_{d1|{\rm in}} (x'-x)} \\
\notag & +
\tilde{{\cal W}}^*_{d1|{\rm out}} \tilde{{\cal R}}_{d1|{\rm out}}
\ep^{{\rm i} q_{d1|{\rm out}} (x'-x)} \\
\notag & +
\Theta(\Omega-\omega) \tilde{{\cal U}}^*_{d2|{\rm in}} 
\tilde{{\cal R}}_{d2|{\rm in}}
\ep^{{\rm i} q_{d2|{\rm in}} (x'-x)} \\
\notag & +
\Theta(\Omega-\omega) \tilde{{\cal U}}^*_{d2|{\rm out}} 
\tilde{{\cal R}}_{d2|{\rm out}}
\ep^{{\rm i} q_{d2|{\rm out}} (x'-x)} \\
\notag & +
\Theta(\Omega-\omega) |S_{d1,d2}|^2 |\tilde{{\cal R}}_{d1|{\rm out}}|^2
\ep^{{\rm i} q_{d1|{\rm out}} (x'-x)} \\
\notag & +
\Theta(\Omega-\omega) (|S_{d2,u}|^2+|S_{d2,d1}|^2) \times \\
\notag & \hphantom{\textrm{}+\textrm{}} |\tilde{{\cal R}}_{d2|{\rm out}}|^2
\ep^{{\rm i} q_{d2|{\rm out}} (x'-x)} \\
& + \rm{c.c.}.
\end{align}
In absence of black hole, the terms involving
coefficients of the $S$-matrix
disappear in (\ref{c06}) and this gives, after integration over
$\omega\in\mathbb{R}^+$, the usual quasi-condensate correlation signal:
$g_{0}^{(2)}(x,x') = n_d F[(x-x')/\xi_d] / \xi_d$.

The non diagonal part is only present if a
horizon exists and only contributes for $\omega<\Omega$; it reads
\begin{align}\label{c07}
\notag \gamma_{0}^{\textrm{non-diag}}(x,x',\omega) &=
\Theta(\Omega-\omega) S^*_{d1,d2} S_{d2,d2} \times \\
\notag & \hphantom{\textrm{}+\textrm{}}
\tilde{\cal R}^*_{d1|\rm{out}} \tilde{\cal R}_{d2|\rm{out}}
\ep^{{\rm i} (q_{d2|{\rm out}}x' - q_{d1|{\rm out}}x)} \\
& + (x \longleftrightarrow x') + \rm{c.c.}.
\end{align}

\subsection{Results for the three configurations}

\begin{figure}
\includegraphics*[width=0.99\linewidth]{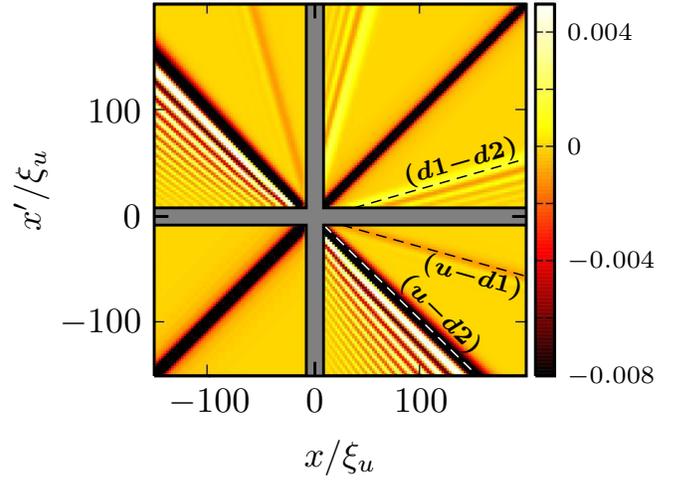}
\caption{(Color online) 2D plot of the numerical result for the
  quantity $\xi_u n_u G^{(2)}_0(x,x')$ in the case of a delta peak
  configuration with $\m_u=0.5$. The shaded area near the axis
  corresponds to the zone $|x|$ or $|x'|<10 \xi_u$. $G^{(2)}_0$ is
  only displayed for $|x|$ and $|x'|>10 \xi_u$, i.e., in the
  asymptotic region where expressions (\ref{c05}), (\ref{c04c}),
  (\ref{c06}) and (\ref{c07}) are valid. The dashed straight lines
  correspond to the correlation lines where a heuristic interpretation
  of the Hawking signal leads to expect the largest signal (see the
  text).}\label{figg2d}
\end{figure}

\begin{figure}
\includegraphics*[width=0.99\linewidth]{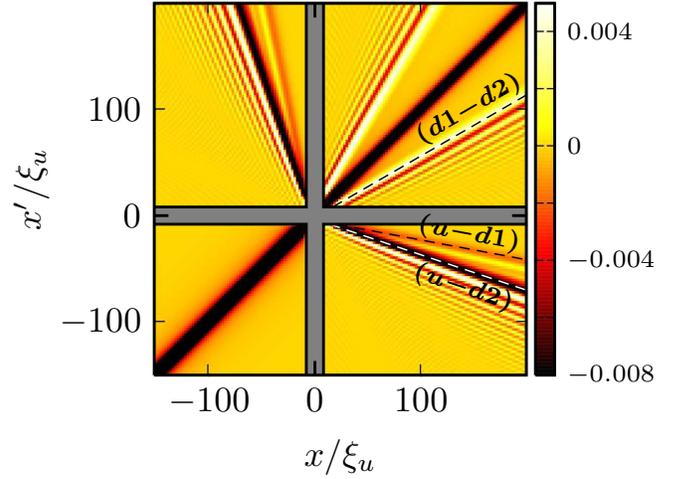}
\caption{(Color online) Same as Fig. \ref{figg2d} for a waterfall
configuration with $\m_u=0.5$.}\label{figg2w}
\end{figure}

Formulae (\ref{c05}), (\ref{c04c}), (\ref{c06}) and (\ref{c07}) allow
us to determine $g^{(2)}_0(x,x')$ through Eq. (\ref{q5}). We performed the
corresponding integration over $\omega\in\mathbb{R}^+$
numerically. The results are shown in Fig. \ref{figg2d} for the delta
peak configuration and in Fig. \ref{figg2w} for the waterfall
configuration. In each of these figures we only display the
correlations for $|x|$ and $|x'|$ larger than a few healing lengths,
because we use formulae which are exact only in the limit $|x|$ and
$|x'|\to\infty$.

In each plot, we also display the lines where the heuristic
interpretation of the Hawking effect presented in the introduction
(see also Ref. \cite{correlations}) leads to locate the more
pronounced correlation signal: if correlated Hawking phonons are
emitted along the $u|{\rm out}$, $d1|{\rm out}$ and $d2|{\rm out}$
channels, at time $t$ after their emission, these phonons are
respectively located at $x_{u|{\rm out}} = (V_u-c_u)t <0$, $x_{d1|{\rm
    out}} = (V_d+c_d)t >0$ and $x_{d2|{\rm out}} = (V_d-c_d)t >0$
\cite{rem2}. This induces a correlation signal along lines of slopes:
$(V_u-c_u)/(V_d+c_d)$ (resulting from $u-d1$ correlations),
$(V_u-c_u)/(V_d-c_d)$ ($u-d2$ correlations) and
$(V_d-c_d)/(V_d+c_d)$ ($d1-d2$ correlations). Of course the lines
with inverse slopes are also present (they correspond to the exchange
$x \leftrightarrow x'$). Indeed, the main features of the
computed $g^{(2)}_0$ perfectly match this interpretation of
the Hawking effect.

These results are very similar to the ones obtained numerically for
the flat profile configuration (already displayed in
Refs. \cite{correlations,Rec09}). This legitimizes the use of density
correlations as a tool for identifying Hawking radiation also in the
realistic delta peak and waterfall configurations.

For each plot the dominant signal is the anti-bunching along the
diagonal ($x=x'$). This corresponds to the typical local density
correlation in a quasi-condensate (cf. Fig.~\ref{fig7}). However, this
signal is modified compared to the one observed in a uniform system
(see, e.g., the discussion in Sec. \ref{outout}). This modification is
connected to non local features which are necessary to verify the sum
rule
\begin{equation}\label{b1bis}
\int_{\mathbb{R}} {\rm d}x' \, g^{(2)}_0(x,x') = \left\{
\begin{array}{lcl}
-n_u & \mbox{when} & x \to -\infty, \\
-n_d & \mbox{when} & x \to +\infty,
\end{array}
\right.
\end{equation}
which is the $T=0$ version of the sum rule (\ref{corr3}) valid when
$x$ is far from the horizon. We checked this sum rule analytically in
Appendix \ref{AppB} on the basis of our Bogoliubov description of the
quantum fluctuations and of the results of Secs. \ref{outout},
\ref{inout} and \ref{inin}.

By comparing Figs. \ref{figg2d} and \ref{figg2w} with the inset of
Fig. \ref{fig7}, one can reverse the argument of Ref. \cite{Fra09} and
argue that (i) in presence of an acoustic horizon, the main new
features of $g^{(2)}_0$ are the non local density correlations which
are simply understood as resulting from the emission of correlated phonons
and (ii) because of the sum rule (\ref{b1bis}), these long range
correlations have to be associated to modifications of the short range
behavior of $g^{(2)}_0$. However, these short range modifications are
not of great experimental relevance because they would be efficiently
blurred by finite temperature effects. The non local aspects instead
are good signatures of the Hawking effect because they are easily
distinguished from the thermal noise (being even reinforced at finite
$T$ as demonstrated in Ref. \cite{Rec09}).

\section{Conclusion and discussion}\label{conclu}

In the present work we have introduced new and realistic acoustic
analogs of black holes and have analyzed the associated Hawking
radiation. We restricted ourself to simple configurations (flat
profile, delta peak, waterfall) but our approach is easily applied to
more complicated cases. For instance, in a wave guide with a
constriction, one is in a mixed situation where there is an external
potential step (as in the waterfall configuration) whereas the non
linear parameter is position-dependent (as in the flat profile
configuration) \cite{Leb03}. This case is in\-te\-res\-ting since it may be
possible to realize it experimentally, but the analytical treatment is
straightforward and we do not consider it here because this would
bring no new insight on our theoretical method. One could also
consider smooth potentials, and the eigen-modes (defined in
Sec. \ref{BdG-equations}) should then be determined numerically, but
the theoretical framework presented here remains of course valid in
this situation. We note also that the present approach can be adapted
to treat the creation of a black hole horizon in a Fermi gas as
suggested in Ref. \cite{Gio05}.

The description of the system in terms of a $S$-matrix allows for a
simple description of the radiation spectrum and a clear
identification of the characteristics of the system. In particular,
we showed that Hawking radiation is absent for a ``no black
hole configuration'' corresponding to a flow connecting two subsonic
asymptotic regions. The spectrum of Hawking radiation has been
computed in the case of a black hole connecting an upstream subsonic
region to a downstream supersonic one, and the concept of Hawking
temperature has been discussed quantitatively.

The main focus has been put on non local density correlations. We
verified that their interpretation in terms of emission of correlated
phonons previously introduced in a model configuration
\cite{correlations,Rec09} also holds in realistic settings.
By studying a sum rule verified by the two-body density matrix, we showed
that the Bogoliubov description of the quantum
fluctuations around the stationary ground state of the system
also provides an accurate description of the short range density
correlations.

In this work we introduced new acoustic black hole configurations
motivated by their possible experimental realization and proposed --
following Ref. \cite{correlations} -- non local density correlations
as practical signatures of Hawking radiation. It is thus important to
discuss if the proposed signal is large enough for being detected
experimentally. From Figs. \ref{figg2d} and \ref{figg2w}, one sees
that the prominent Hawking signal corresponds to $u-d2$
correlations. For each figure this corresponds to a line of negative
correlation where the largest value of $\xi_u n_u G^{(2)}_0$ is
between $-0.01$ and $-0.02$. In present-day experiments it is
possible to measure density fluctuations around a mean density $n_u$
of order 10 $\mu$m$^{-1}$ in a setting where the healing length
$\xi_u$ is around a few $\mu$m (see, e.g., Ref. \cite{Arm11}). It is
thus realistic to hope to reach a configuration where
$|G^{(2)}_0|_{\rm max} \simeq 5 \times 10^{-3}$, and for detecting a
signal of this intensity a precision of around $10^{-4}$ is required
on the determination of $G^{(2)}$.  Noticing that $G^{(2)}$ is the
{\it quadratic} relative density fluctuations, detecting this signal
would correspond to measuring density fluctuations with a precision of
order of 1 \%, which seems within reach of present-day experimental
techniques.

\begin{acknowledgments}

We thank V. E. Kravtsov and C. I. Westbrook for fruitful discussions.
This work was supported by the IFRAF Institute, by
Grant ANR-08-BLAN-0165-01 and by ERC through the QGBE grant.
A. R. acknowledges the kind hospitality of the LPTMS in Orsay.

\end{acknowledgments}

\appendix

\section{Low energy behavior of the
scattering matrix}\label{AppA}

In this appendix we display the analytical results for the
low-$\omega$ behavior of combinations of the elements of the
$S$-matrix relevant for computations of the Hawking temperature and
for the fulfillment of the sum rule (\ref{b1bis}). We only give
results for the flat profile and waterfall configurations because
those concerning the delta peak configuration are too long. Indeed,
for the delta peak configuration, a numerical determination of the
coefficients of the $S$-matrix (which simply amounts to invert a $4
\times 4$ matrix) is more convenient than the analytical approach. We
checked that both agreed to an extremely good accuracy.

\subsection{Flat profile configuration}

In this case, the scattering coefficients depend of the two Mach
numbers $\m_u$ and $\m_d$. The coefficients $f$ and $h$ defined in
Eqs. (\ref{loww}) verify the following relations

\begin{equation}
\label{Fud2ud2 - Flat profile configuration}
|f_{u,d2}|^2=2\frac{\mathpzc{m}_{u}}{\mathpzc{m}_{d}}
\frac{\mathpzc{m}_{u}^{2}}{\mathpzc{m}_{d}^{2}-\mathpzc{m}_{u}^{2}}
\frac{1-\mathpzc{m}_{u}}{1+\mathpzc{m}_{u}}
(\mathpzc{m}_{d}^{2}-1)^{\frac{3}{2}},
\end{equation}

\begin{equation}
\label{Fd1d2d1d2 - Flat profile configuration}
|f_{d1,d2}|^2=\frac{1}{2}
\left(\frac{\mathpzc{m}_{u}}{\mathpzc{m}_{d}}\right)^{2}
\frac{\mathpzc{m}_{d}-\mathpzc{m}_{u}}{\mathpzc{m}_{d}+\mathpzc{m}_{u}}
\frac{1-\mathpzc{m}_{u}}{1+\mathpzc{m}_{u}}
(\mathpzc{m}_{d}^{2}-1)^{\frac{3}{2}},
\end{equation}

\begin{equation}
\label{Fd2d2d2d2 - Flat profile configuration}
|f_{d2,d2}|^2=\frac{1}{2}
\left(\frac{\mathpzc{m}_{u}}{\mathpzc{m}_{d}}\right)^{2}
\frac{\mathpzc{m}_{d}+\mathpzc{m}_{u}}{\mathpzc{m}_{d}-\mathpzc{m}_{u}}
\frac{1-\mathpzc{m}_{u}}{1+\mathpzc{m}_{u}}
(\mathpzc{m}_{d}^{2}-1)^{\frac{3}{2}}, 
\end{equation}

\begin{equation}
\label{Fud2d1d2 - Flat profile configuration}
f_{u,d2}^*f_{d1,d2}=
\left(\frac{\mathpzc{m}_{u}}{\mathpzc{m}_{d}}\right)^{\frac{3}{2}}
\frac{\mathpzc{m}_{u}}{\mathpzc{m}_{d}+\mathpzc{m}_{u}}
\frac{1-\mathpzc{m}_{u}}{1+\mathpzc{m}_{u}}
(\mathpzc{m}_{d}^{2}-1)^{\frac{3}{2}},
\end{equation}

\begin{equation}
\label{Fud2d2d2 - Flat profile configuration}
f_{u,d2}^*f_{d2,d2}
=-\left(\frac{\mathpzc{m}_{u}}{\mathpzc{m}_{d}}\right)^{\frac{3}{2}}
\frac{\mathpzc{m}_{u}}{\mathpzc{m}_{d}-\mathpzc{m}_{u}}
\frac{1-\mathpzc{m}_{u}}{1+\mathpzc{m}_{u}}
(\mathpzc{m}_{d}^{2}-1)^{\frac{3}{2}},
\end{equation}

\begin{equation}
\label{Fd1d2d2d2 - Flat profile configuration}
f_{d1,d2}^*f_{d2,d2}=-\frac{1}{2}
\left(\frac{\mathpzc{m}_{u}}{\mathpzc{m}_{d}}\right)^{2}
\frac{1-\mathpzc{m}_{u}}{1+\mathpzc{m}_{u}}
(\mathpzc{m}_{d}^{2}-1)^{\frac{3}{2}},
\end{equation}

\begin{equation}
\label{Gud2ud2 - Flat profile configuration}
\mbox{Re}(f^*_{u,d2}h_{u,d2})
=-\frac{\mathpzc{m}_{u}}{(1+\mathpzc{m}_{u})^{2}}.
\end{equation}

\subsection{Waterfall configuration}

In the case of the waterfall configuration, the elements of the $S$-matrix
depend on a unique parameter; we chose to express them as functions
of the Mach number $\mathpzc{m}_{u}$.
\begin{equation}
\label{Fud2ud2 - Waterfall configuration}
|f_{u,d2}|^2=2
\frac{\mathpzc{m}_{u}(1-\mathpzc{m}_{u})^{\frac{3}{2}}
(1+\mathpzc{m}_{u}^{2})^{\frac{3}{2}}}{(1+\mathpzc{m}_{u})^{\frac{1}{2}}
(1+\mathpzc{m}_{u}+\mathpzc{m}_{u}^{2})^{2}},
\end{equation}

\begin{equation}
\label{Fd1d2d1d2 - Waterfall configuration}
|f_{d1,d2}|^2=\frac{1}{2}
\frac{(1-\mathpzc{m}_{u})^{\frac{7}{2}}
(1+\mathpzc{m}_{u}^{2})^{\frac{3}{2}}}{(1+\mathpzc{m}_{u})^{\frac{1}{2}}
(1+\mathpzc{m}_{u}+\mathpzc{m}_{u}^{2})^{2}},
\end{equation}

\begin{equation}
\label{Fd2d2d2d2 - Waterfall configuration}
|f_{d2,d2}|^2=\frac{1}{2}\frac{(1-\mathpzc{m}_{u}^{4})^{\frac{3}{2}}}
{(1+\mathpzc{m}_{u}+\mathpzc{m}_{u}^{2})^{2}},
\end{equation}

\begin{equation}
\label{Fud2d1d2 - Waterfall configuration}
f_{u,d2}^*f_{d1,d2}=
-\frac{\mathpzc{m}_{u}^{\frac{1}{2}}(1-\mathpzc{m}_{u})^{\frac{5}{2}}
(1+\mathpzc{m}_{u}^{2})^{\frac{3}{2}}}{(1+\mathpzc{m}_{u})^{\frac{1}{2}}
(1+\mathpzc{m}_{u}+\mathpzc{m}_{u}^{2})^{2}},
\end{equation}

\begin{equation}
\label{Fud2d2d2 - Waterfall configuration}
f_{u,d2}^*f_{d2,d2}=-\frac{\mathpzc{m}_{u}^{\frac{1}{2}}
(1+\mathpzc{m}_{u})^{\frac{1}{2}}(1-\mathpzc{m}_{u})^{\frac{3}{2}}
(1+\mathpzc{m}_{u}^{2})^{\frac{3}{2}}}
{(1+\mathpzc{m}_{u}+\mathpzc{m}_{u}^{2})^{2}},
\end{equation}

\begin{equation}
\label{Fd1d2d2d2 - Waterfall configuration}
f_{d1,d2}^*f_{d2,d2}=\frac{1}{2}
\frac{(1+\mathpzc{m}_{u})^{\frac{1}{2}}(1-\mathpzc{m}_{u})^{\frac{5}{2}}
(1+\mathpzc{m}_{u}^{2})^{\frac{3}{2}}}{(1+\mathpzc{m}_{u}+\mathpzc{m}_{u}^{2})^{2}},
\end{equation}

\begin{equation}
\label{Gud2ud2 - Waterfall configuration}
\mbox{Re}(f_{u,d2}^*h_{u,d2})
=-\frac{\mathpzc{m}_{u}(2+\mathpzc{m}_{u}^{2})
(1+2\mathpzc{m}_{u}^{2})}{(1+\mathpzc{m}_{u})^{2}
(1+\mathpzc{m}_{u}+\mathpzc{m}_{u}^{2})^{2}}.
\end{equation}

\section{Computation of the energy current}\label{AppC}

In this appendix we briefly indicate how formula (\ref{rs13}) is
obtained from Eqs. (\ref{rs4}) and (\ref{rs5}). For a point $x$ deep
in the subsonic region, using the $\eta$-unitarity (\ref{e19}) of the
$S$-matrix, one can write the relevant contributions to
$J_0(x,\omega)$ in (\ref{rs5}) under the form
\begin{align}\label{c1}
\notag J_0(x,\omega) &= (q_{u|{\rm in}} - k_u) |\tilde{{\cal W}}_{u|{\rm in}}|^2
+ (q_{u|{\rm out}} - k_u) |\tilde{{\cal W}}_{u|{\rm out}}|^2 \\
\notag &+ |S_{u,d2}|^2 \Big[
(q_{u|{\rm out}} + k_u) |\tilde{{\cal U}}_{u|{\rm out}}|^2 \\
&\hphantom{+ |S_{u,d2}|^2 \Big[} + 
(q_{u|{\rm out}} - k_u) |\tilde{{\cal W}}_{u|{\rm out}}|^2 \Big].
\end{align}
A simple but lengthly computation shows that the contributions to
$\Pi_0$ of the two first terms of the r.h.s. of (\ref{c1}) cancel
after integration over $\omega$. This is very satisfactory because
this shows that there is no Hawking radiation when $S_{u,d2}=0$, i.e.,
in absence of black hole.

Using Eq. (\ref{cc}) the remaining can be written as $|S_{u,d2}|^2
J_{u|{\rm out}}/\xi_u c_u$, and since $\hbar/(m c_u \xi_u)=1$ this
directly yields Eq. (\ref{rs13}). The minus sign in this formula comes
from the fact that $J_{u|{\rm out}}=-1$ and corresponds to the
direction of propagation of the energy in the $u|{\rm out}$ mode.

\section{Verification of the sum rule (\ref{corr3})}\label{AppB}

In this appendix we check that the Bogoliubov approach based on expansion
(\ref{q1}) indeed makes it possible
to verify Eq. (\ref{b1bis}) which is
the $T=0$ version of the sum rule (\ref{corr3}) when $x$ is
far from the sonic horizon, either upstream or downstream.

We start here by a technical remark. For fixed $x$, the contributions
to $\gamma_0(x,x',\omega)$ displayed in Secs. \ref{outout}, \ref{inout}
and \ref{inin} are only noticeable for $x'\ll-\xi_u$ or
$x'\gg\xi_d$. As a result, the analytical forms displayed in these
sections can be extended for all $x'\in\mathbb{R}$ without introducing
noticeable errors in the computation of the integral of
$\gamma_0(x,x',\omega)$ over $x'$. Then, the integration of
$\gamma_0(x,x',\omega)$ over $x'$ just amounts to evaluating the
following integral:
\begin{align}\label{b0}
& \int_\mathbb{R} {\rm d}x' \, \ep^{{\rm i} q_\ell(\omega) x'} = \\
\notag & \left\{
\begin{array}{ccl}
0 & \mbox{if} & \ell \in \{d1|{\rm in},d2|{\rm in}\}, \\
|V_g(Q_\ell(0))| \delta(\omega) & \mbox{if} & \ell \in
\{u|{\rm in},u|{\rm out},d1|{\rm out},d2|{\rm out}\}.
\end{array}
\right.
\end{align}
In (\ref{b0}) we used the fact that $q_{d1|{\rm in}}$ and $q_{d2|{\rm
in}}$ never cancel, whereas the other $q_\ell$'s do for
$\omega=0$ (see Fig. \ref{fig3}). Using this prescription, for $x$
large and negative, one gets from Eq. (\ref{c05})
\begin{align}\label{b2}
\notag & \int_{\mathbb{R}^-} {\rm d}x' \, n_u
\int_0^\infty \frac{{\rm d}\omega}{2 \pi} \,
\gamma_0(x,x',\omega) \\
& = -n_u +
\frac{n_u}{2} \frac{|f_{u,d2}|^2}{1-\m_u},
\end{align}
and from Eq. (\ref{c04c})
\begin{align}\label{b3}
\notag & \int_{\mathbb{R}^+} {\rm d}x' \, \sqrt{n_u n_d}
\int_0^\infty \frac{{\rm d}\omega}{2 \pi} \,
\gamma_0(x,x',\omega) \\
& = \frac{n_u}{2}
\sqrt{\frac{c_u}{c_d}\frac{n_d}{n_u}} \, \mbox{Re}
\left[
\frac{f_{u,d2}^*}{1-\m_u}(f_{d1,d2}+f_{d2,d2})\right]. 
\end{align}
Altogether this yields
\begin{align}\label{b4}
\notag & \lim_{x\to-\infty} \int_{\mathbb{R}} {\rm d}x' \, 
g_0(x,x') = -n_u \\
& +
\frac{n_u}{2}
\sqrt{\frac{c_u}{c_d} \frac{n_d}{n_u}} \, \mbox{Re}
\left(\frac{f_{u,d2}^*}{1-\m_u} {\cal F}\right),
\end{align}
where
\begin{equation}\label{b5}
{\cal F} = f_{u,d2}
\sqrt{\frac{c_d}{c_u} \frac{n_u}{n_d}} + f_{d1,d2} + f_{d2,d2}.
\end{equation}
Similarly one gets [from Eqs. (\ref{c04c}), (\ref{c06}) and (\ref{c07})]
\begin{align}\label{b6}
\notag & \lim_{x\to+\infty} \int_{\mathbb{R}} {\rm d}x' \, g^{(2)}_0(x,x') = -n_d \\
& +
\frac{n_d}{2}
\left(\frac{c_u}{c_d}\right)^2 \mbox{Re}
\left[
\left(\frac{f_{d1,d2}^*}{\m_d+1}+\frac{f_{d2,d2}^*}{\m_d-1}\right)
{\cal F}\right].
\end{align}
Using the analytical expressions for the combinations of coefficients
$f_{u,d2}$, $f_{d1,d2}$ and $f_{d2,d2}$ displayed in Appendix
\ref{AppA}, one can easily verify that the second terms of the r.h.s.
of Eqs. (\ref{b4}) and (\ref{b6}) cancel.  This is due to
the fact that ${\cal F}$ is identically null in the flat profile and
waterfall configurations (this is more tedious to check but we
confirmed it analytically). The same holds for the delta peak
configuration. This shows that the sum rule (\ref{b1bis}) is fulfilled
in these three cases. This is a strong confirmation of both the
validity of the Bogoliubov approach and of the exactness of our
analytical results.


\begin{thebibliography}{99}
\bibitem{remarque} Remember that the value of $\beta_d$
  depends on the configuration considered, for instance $\beta_d=0$
  in the flat profile configuration.
\bibitem{qu-optics} For our study it is more appropriate to withdraw
  the disconnected terms in (\ref{q4}) and we thus do not follow the
  usual quantum optics notation.
\bibitem{rem1} In this case one does not have the complicated
  scattering modes (\ref{e17}) relevant for a black hole
  configuration. One just has two free modes -- propagating either to
  the left or to the right -- which are exactly identical to the modes
  $u|{\rm in}$ and $u|{\rm out}$ identified in Sec. \ref{general}, and
  the $S$-matrix is the $2\times 2$ identity.
\bibitem{misprints} In Eqs. (27), (29), (33), (36), (39) and (40) of
  Ref. \cite{Rec09} the prefactor should be $1/(2\pi)$ instead of
  $1/(4\pi)$. Also the argument of the exponential in formulae (27)
  and (33) has the wrong sign.
\bibitem{rem2} This is obtained assuming linear dispersion relations
  along each branch of the dispersion relation, which is a long wavelength 
approximation.
\end{thebibliography}
\end{document}